\RequirePackage[utf8]{inputenc}
\documentclass[]{imag-ms-template}

\usepackage[english]{babel}
\usepackage[backend=biber, style=apa]{biblatex}
\usepackage{pdflscape}

\usepackage{csquotes}
\usepackage{graphicx}
\usepackage{subcaption}
\usepackage{float}
\usepackage{hyperref}
\usepackage{amsmath, amssymb, amsthm}
\usepackage{booktabs}
\usepackage{breqn}
\usepackage{authblk}
\usepackage{xr}
\usepackage{cleveref}
\usepackage[english]{babel}
\usepackage[backend=biber, style=apa]{biblatex}
\usepackage{csquotes}
\usepackage{graphicx}
\usepackage{subcaption}
\usepackage{float}
\usepackage{hyperref}
\usepackage{amsmath, amssymb, amsthm}
\usepackage{booktabs}
\usepackage{breqn}
\usepackage{authblk}
\usepackage{pdflscape}
\usepackage{placeins}
\usepackage{lmodern}
\usepackage[T1]{fontenc}
\usepackage{subfiles}
\newif\ifarxiv

\arxivtrue

\ifarxiv
  \expandafter\def\csname suppfallback@app:feature_extr\endcsname{S1}
\expandafter\def\csname suppfallback@app:baselines\endcsname{S2}
\expandafter\def\csname suppfallback@app:add_rel_work\endcsname{S3.1}
\expandafter\def\csname suppfallback@app:hyperlitreview\endcsname{S3.2}
\expandafter\def\csname suppfallback@tab:hyperparameter_tuning_simple\endcsname{S1}
\expandafter\def\csname suppfallback@tab:hyperparameter_tuning_transposed\endcsname{S2}
\expandafter\def\csname suppfallback@tab:transposed_hyperparams_extended\endcsname{S3}
\expandafter\def\csname suppfallback@tab:hyperparameter_tuning_transposed_short\endcsname{S4}
\expandafter\def\csname suppfallback@tab:detailed_hyperparameter_comparison\endcsname{S5}
\expandafter\def\csname suppfallback@app:encoder_comp\endcsname{S3.3}
\expandafter\def\csname suppfallback@fig:encodercomp\endcsname{S1}
\expandafter\def\csname suppfallback@app:hyperparametertuning\endcsname{S3.4}
\expandafter\def\csname suppfallback@appfig:modelcomphyper\endcsname{S2}
\expandafter\def\csname suppfallback@tab:hyperparams\endcsname{S6}
\expandafter\def\csname suppfallback@fig:hypercomp\endcsname{S3}
\expandafter\def\csname suppfallback@app:group_vs_indiv\endcsname{S3.5}
\expandafter\def\csname suppfallback@fig:groupvsindiv\endcsname{S4}
\expandafter\def\csname suppfallback@app:noiseceil\endcsname{S4}
\expandafter\def\csname suppfallback@tab:significance\endcsname{S7}
\expandafter\def\csname suppfallback@fig:loss_afm_sfm\endcsname{S5}
\expandafter\def\csname suppfallback@fig:uncertaintyseries\endcsname{S6}
\expandafter\def\csname suppfallback@app:flatmaps\endcsname{S5.4}
\expandafter\def\csname suppfallback@fig:yeonetworks\endcsname{S7}
\expandafter\def\csname suppfallback@fig:trueactivation\endcsname{S8A}
\expandafter\def\csname suppfallback@sub@fig:trueactivation\endcsname{A}
\expandafter\def\csname suppfallback@fig:noiseceilmap\endcsname{S8B}
\expandafter\def\csname suppfallback@sub@fig:noiseceilmap\endcsname{B}
\expandafter\def\csname suppfallback@fig:noiseceilingparcels\endcsname{S8}
\expandafter\def\csname suppfallback@fig:afmparcelmap\endcsname{S9}
\expandafter\def\csname suppfallback@fig:glmparcelmap\endcsname{S10A}
\expandafter\def\csname suppfallback@sub@fig:glmparcelmap\endcsname{A}
\expandafter\def\csname suppfallback@fig:sfmparcelmap\endcsname{S10B}
\expandafter\def\csname suppfallback@sub@fig:sfmparcelmap\endcsname{B}
\expandafter\def\csname suppfallback@fig:glm_sfm_flatmaps\endcsname{S10}
\expandafter\def\csname suppfallback@fig:afmvsglmflatmap\endcsname{S11A}
\expandafter\def\csname suppfallback@sub@fig:afmvsglmflatmap\endcsname{A}
\expandafter\def\csname suppfallback@fig:afmvssfmflatmap\endcsname{S11B}
\expandafter\def\csname suppfallback@sub@fig:afmvssfmflatmap\endcsname{B}
\expandafter\def\csname suppfallback@tab:noiseceiling\endcsname{S8}
\expandafter\def\csname suppfallback@tab:noiseceilingnetwork\endcsname{S9}
\expandafter\def\csname suppfallback@fig:crps_networks\endcsname{S12}
\expandafter\def\csname suppfallback@fig:crps_diff_flatmap\endcsname{S13}
\expandafter\def\csname suppfallback@fig:contwindow_networks\endcsname{S14}
\expandafter\def\csname suppfallback@fig:predwindow_networks\endcsname{S15}

  \newcommand{\suppfigref}[1]{Supplementary Figure~\csname suppfallback@#1\endcsname}
  \newcommand{\supptabref}[1]{Supplementary Table~\csname suppfallback@#1\endcsname}
  \newcommand{\suppsecref}[1]{Supplementary Section~\csname suppfallback@#1\endcsname}
 
  \newcommand{\suppseconlyref}[1]{Section~\csname suppfallback@#1\endcsname}
  \newcommand{\suppnumref}[1]{\csname suppfallback@#1\endcsname}
\else
  \usepackage{xr-hyper}
  \newcommand{\suppfigref}[1]{Supplementary Figure~\ref{#1}}
  \newcommand{\supptabref}[1]{Supplementary Table~\ref{#1}}
  \newcommand{\suppsecref}[1]{Supplementary Section~\ref{#1}}
  \newcommand{\suppseconlyref}[1]{Section~\ref{#1}}
  \newcommand{\suppnumref}[1]{\ref{#1}}

\fi

\begin{document}
\title{Probabilistic Prediction of Neural Dynamics via Autoregressive Flow Matching}
\author[1]{Nicole Rogalla}
\author[1]{Yuzhen Qin}
\author[2, 3]{Mario Senden}
\author[1]{Ahmed El-Gazzar}
\author[1]{Marcel van Gerven}
\affil[1]{Department of Machine Learning and Neural Computing, Donders Institute for Brain, Cognition and Behaviour, Radboud University, Nijmegen, the Netherlands}
\affil[2]{Department of Cognitive Neuroscience, Maastricht University, Maastricht, the Netherlands}
\affil[3]{Maastricht Brain Imaging Center, Maastricht University, Maastricht, the Netherlands}
\date{}
\maketitle
\vspace{-3em}
\begin{abstract}
Forecasting neural activity in response to naturalistic stimuli remains a key challenge for understanding brain dynamics and enabling downstream neurotechnological applications. Here, we introduce a generative forecasting framework for modeling neural dynamics based on autoregressive flow matching (AFM). Building on recent advances in transport-based generative modeling, our approach probabilistically predicts neural responses at scale from multimodal sensory input. Specifically, we learn the conditional distribution of future neural activity given past neural dynamics and concurrent sensory input, explicitly modeling neural activity as a temporally evolving process in which future states depend on recent neural history. 
We evaluate our framework on the Algonauts project 2025 challenge functional magnetic resonance imaging dataset using subject-specific models. AFM significantly outperforms both a non-autoregressive flow-matching baseline and the official challenge general linear model baseline in predicting short-term parcel-wise blood oxygenation level-dependent (BOLD) activity, demonstrating improved generalization and widespread cortical prediction performance. Ablation analyses show that access to past BOLD dynamics is a dominant driver of performance, while autoregressive factorization yields consistent, modest gains under short-horizon, context-rich conditions. Together, these findings position autoregressive flow-based generative modeling as an effective approach for short-term probabilistic forecasting of neural dynamics with promising applications in closed-loop neurotechnology. 

\end{abstract}

\section{Introduction}

Predicting how brain activity evolves over time in response to the environment is central to understanding neural systems and represents a key step toward closed-loop neurotechnology. Forecasting models provide a principled framework for investigating neural information processing, evaluating computational theories, serving as a key step toward closed-loop neurotechnology, where predicted neural activity informs adaptive, personalized interventions~\parencite{Xiong2023}. The challenge of forecasting neural dynamics is particularly pronounced in naturalistic settings, where complex, dynamic stimuli unfold over time, increasing predictive complexity while providing greater ecological validity~\parencite{zhou2025realworld}.

Across neural recording modalities, forecasting has emerged as a key focus to modeling neural dynamics~\parencites{antoniades2024neuroformer, immer2025, lueckmann2025zapbench,duan2025, Li2023, lu2025, chehab2022}. In functional magnetic resonance imaging (fMRI), predictive approaches largely rely on encoding models that predict neural activity based on stimulus inputs~\parencite{Naselaris2011}, with examples including~\textcite{Khosla2021,Guclu2017, Guclu10005, Yamins2016, eren2025, scholz2025}.
The recent Algonauts project 2025 challenge further highlights the growing interest in predictive modeling under ecologically valid conditions by calling for the development of fMRI encoding models for multi-modal naturalistic stimuli~\parencite{gifford2025}. Building on the challenge-winning encoding model TRIBE \parencite{tribe2025}, recent work has extended this approach into a large-scale foundation model for in-silico neuroscience~\parencite{dAscoli2026TribeV2}.
Parallel advances in blood oxygenation level-dependent (BOLD) dynamics modeling have shown strong predictive performance in resting-state fMRI~\parencites{sun2024, caro2024brainlm, Sobczak2021, Dvornek2019, Wein2022}. 
Yet, few dynamics approaches extend to task-based fMRI modeling~\parencites{Paugam2024, Dorin2024}, with even fewer integrating both past stimuli and neural dynamics to forecast future BOLD responses, despite evidence of endogenous activity influencing stimulus processing~\parencites{Amos1996, Dehaghani2025PreStimulus}. 

Despite these advances, most forecasting approaches remain deterministic, disregarding the intrinsic variability of neural dynamics and the noisy and indirect nature of fMRI recordings~\parencites{LIU2016141, Faisal2008Noise, TOMKO1974405}. This limitation is particularly relevant in naturalistic experimental settings, where participants are exposed to continuous, high-dimensional real-world stimuli with increased complexity~\parencites{SIMONY2020116461,ZHANG2021100298, Gong2023, gifford2025}. 
Generative models address this issue by yielding predictive distributions over possible future trajectories. This perspective resonates with the view that the brain itself may operate probabilistically: integrating noisy sensory inputs, internal states, and prior beliefs to generate behavior and perception aiming to reduce the prediction error~\parencite{millidge2022}. Generative forecasting frameworks therefore not only improve variability modeling but also provide a principled connection to theories of neural computation. 

Probabilistic models of fMRI data have primarily been employed for decoding of external variables, such as movie ratings, from neural data, hemodynamic or connectivity modeling~\parencites{Battle2006, Svensen2000, AlowadiShenTino2016, FRISTON2003, HARRISON2015217, LiTao2011BrainActivities, Ajith2024, SafariMohammadbeigi2012, FRISTON2003}. Apart from recent exceptions such as a hierarchical diffusion model for fMRI forecasting introduced by \textcite{hu2025synthesizing}, and the Brain Foundation Model~\parencite{bayazi2024generalpurpose}, which builds on the success of large language models (LLMs), probabilistic models focusing on fMRI dynamics forecasting remain rare, leaving a clear methodological gap.

Beyond fMRI, probabilistic deep generative models have demonstrated strong performance in time series forecasting across diverse domains~\parencites{Kollovieh2023, rasul2021,meijer2024risediffusionmodelstimeseries,Feng_Miao_Zhang_Zhao_2024}. However, their computational costs have motivated the development of flow matching (FM) as a new training paradigm~\parencite{lipman2023}. FM enables stable and efficient training by learning the transformation between simple base distribution and target distribution without the need for simulation. FM has recently been applied to time series forecasting~\parencites{kollovieh2025, hu2025, Zhang2024} with strong empirical results and fast training and sampling. 

Existing FM methods, however, typically model the conditional distribution of the entire future trajectory, which poses a difficult optimization problem. To address this, \textcite{elgazzar2025AFM} proposed \textit{autoregressive flow matching} (AFM), which decomposes forecasting into a sequence of one-step conditional distributions and requires autoregressively generating the next-step prediction during sampling. Their empirical results across classical dynamical systems and real-world data benchmarks show that AFM improves predictive accuracy compared to non-autoregressive FM baselines. 
This autoregressive formulation aligns the generative process with the temporal structure of the data. AFM is a natural fit for fMRI forecasting, as it unites the computational efficiency of flow-based generative models with the autoregressive modeling principles traditionally employed in fMRI modeling through Granger causality analyses and vector autoregressive models~\parencites{ROEBROECK2005, Deshpande2009, GARG2011}. By explicitly modeling temporal dependencies between successive states within the future segment, this approach further resonates with a dynamical systems perspective that interprets brain activity as an evolving process over time~\parencite{Favela2021}.

In this work, we introduce a large-scale generative framework for modeling neural dynamics using autoregressive flow matching that predicts parcel-wise BOLD activity in the short-term future. By learning the conditional distribution of future neural activity given both past BOLD dynamics and naturalistic sensory input, our model moves beyond stimulus-only and deterministic approaches, enabling uncertainty-aware predictions and realistic generative sampling of neural trajectories.
This framework combines the neuroscientific appeal of autoregressive modeling with the stability and efficiency of modern flow-based generative methods.
Using the large-scale Algonauts project 2025 challenge fMRI dataset, we demonstrate predictive performance and uncertainty quantification of neural activity at scale. AFM significantly outperforms both the official challenge general linear model baseline and a non-autoregressive FM baseline in forecasting multi-step short-term parcel-wise BOLD activity, yielding consistently higher noise-ceiling–adjusted correlations across subjects and cortical regions. These results indicate that AFM approaches the performance noise ceiling imposed by the measurement modality, which is the maximum predictable variance given measurement noise and intrinsic neural variability, bringing neural forecasting closer to the accuracy required for downstream applications.

\section{Methods}\label{chap:methods}

\subsection{Dataset}

This study utilizes datasets form the Courtois Project on Neural Data Modeling (NeuroMod, \url{https://www.cneuromod.ca/}) \textit{friends} and \textit{movie10} as provided by the Algonauts project 2025 challenge "How the Human Brain Makes Sense of Multimodal Movies"~\parencite{gifford2025, Boyle2023}. It contains single-subject BOLD response recordings to naturalistic stimuli for four subjects. 
FMRI BOLD responses were recorded using a 3T Siemens Prisma Fit scanner with a repetition time (TR) of 1.49s. Responses were normalized to the Montreal Neurological Institute (MNI) spatial template and averaged to 1000 functionally defined brain parcels~\parencite{Schaefer2017}. 
The \textit{friends} dataset encompasses seasons 1 to 6 of the Sitcom \textit{Friends}. The \textit{movie10} dataset consists of three feature movies (\textit{Bourne Supremacy}, \textit{The Wolf of Wall Street}, \textit{Hidden Figures}) and a documentary (\textit{Life documentary}), with textit{Hidden Figures} and \textit{Life} having been watched twice. 
The dataset encompassed approximately 65 hours (55 hours of \textit{friends} and 10 hours of \textit{movie10}) of movie stimuli content.
Stimuli were presented in English and are available as movie frames, audio samples and language transcripts. 
For more detailed information on the fMRI dataset, please consult the Courtois Neuromod's Project documentation (\url{https://docs.cneuromod.ca}).

\subsection{Relation to the Algonauts project 2025 challenge}

The Algonauts project 2025 challenge focuses on stimulus-only encoding models and explicitly prohibits the use of past BOLD dynamics as inputs to the prediction model. The phase 1 leaderboard evaluates predictions for \textit{Friends} season 7, whose BOLD responses have not been publicly released. 
Our approach explicitly utilizes past BOLD dynamics to model temporal dependencies. As a result, our framework cannot be applied to the official season 7 test set and we instead evaluate forecasting performance on season 6. Our results are therefore not directly comparable to the challenge leaderboard due to differences in training and test set, model input (stimulus and BOLD dynamics vs. stimulus-only) and difference in prediction horizon, given that we perform multi-step forecasting while challenge models are focused on one-step prediction. 
We report the official general linear model (GLM) baseline as a classical encoding model reference that anchor our results within the challenge context. Retraining winning challenge models under our settings was not pursued due to the substantial computational demands of the top-performing approach~\parencite{tribe2025}. 

\subsection{Stimulus features}
Stimulus features were extracted using the Algonauts project 2025 challenge development kit~\parencite{gifford2025}, comprising visual, auditory and linguistic embeddings from pretrained models. Visual features were extracted using a Slow R50~\parencite{Feichtenhofer_2019_ICCV} pretrained on Kinetics 400~\parencite{kay2017}, capturing spatiotemporal action information.
Auditory features were mel-frequency cepstral coefficients representing human frequency perception~\parencite{Abdul2022}.
Linguistic features were obtained from the large language model \textit{Bidirectional Encoder Representations from Transformers} (BERT)~\parencite{devlin2019}, providing contextual word embeddings from the episode transcripts. More details on feature extraction can be found in \suppsecref{app:feature_extr}.

\subsection{Flow-matching-based generative modeling}
Before describing our autoregressive flow matching framework, we briefly outline the principles of flow-based generative modeling that underpin our approach.
Generative models aim to learn a mapping from a simple base distribution $p^0$ to a complex data distribution $p^1$~\parencite{Ruthotto2021}. This transformation can be defined through a continuous normalizing flow (CNF) as a solution to an ordinary differential equation (ODE)~\parencite{Chen2019}. This flow $\psi \colon \Omega \times [0,1] \to \Omega$ on data space $\Omega$ evolves a sample $x^0 \sim p^0$ into a target sample $x^1\coloneq \psi(x^0,1) \sim p^1$ by solving the initial value problem 
\begin{equation}
\mathrm{d}\psi(x, s) = \mu(\psi(x, s), s)\,\mathrm{d}s, 
\quad \psi(x, 0) = x_0, 
\quad s \in [0, 1] .
\end{equation}
The vector field $\mu \colon \Omega \times [0,1] \to \Omega$, that defines the velocity of the flow and generates the probability path $(p^s)^{0\leq s \leq 1}$, is to be learned.

Conditional flow matching~\parencite{lipman2023} is a simulation-free training method of CNFs that avoids numerically solving ODEs during training. Instead, it learns a neural vector field $v_\theta$ with trainable parameters $\theta$ to approximate the intractable target vector field $\mu$ by regressing against per-example conditional optimal vector fields $\mu({x}, s | z)$ conditioned on an arbitrary random variable $z$ with probability density function $p_z$:
\begin{equation}
\mathcal{L}_{\text{CFM}}(\theta) = \mathbb{E}_{s\sim U(0,1), z\sim p_z, x\sim p^s({x}|z)} \| \mu({x}, s | z) -v_\theta({x},s) \|^2 .
\end{equation}
This framework allows efficient and stable training of generative models and provides the basis for our autoregressive flow matching extension.

\subsection{Autoregressive flow matching}\label{chap:frameworks}

\paragraph{Problem setting}

Let ${y}_\tau \in \mathbb{R}^n$ denote the fMRI BOLD response of $n$ brain parcels at time $\tau$. Given access to $l$ past fMRI images including the current image, denoted as ${Y}_l = \{{y}_{\tau - l}, \ldots,{y}_\tau\}$, the objective is to predict the next $f$ future images ${Y}_f = \{{y}_{\tau + 1}, \ldots, {y}_{\tau + f}\}$, conditioned on both past images and stimulus feature input ${C} = \{{C}_l, {C}_f\}$ with past stimulus features ${C}_l= \{{c}_{\tau - l}, \ldots, {c}_{\tau}\}$ and future stimulus features $ {C}_f = \{{c}_{\tau +1}, \ldots, {c}_{\tau + f}\}$. 
Formally, the goal is to learn a conditional distribution $p({Y}_f \mid {Y}_l, {C})$ from a dataset $\mathcal{D} = \left\{ ({Y}_f^i, {Y}_l^i, {C}^i) \right\}_{i=1}^m$ composed of $m$ training sequences. Once learned, this distribution can be sampled to generate future BOLD trajectories conditioned on observed history and stimulus features.

\paragraph{Training approach}

AFM introduced by \textcite{elgazzar2025AFM} is the backbone of our generative modeling framework. BOLD time series forecasting is treated as sampling from a learned conditional distribution of future BOLD images given historical images and stimulus features. Unlike standard flow matching, which predicts the all $f$ future BOLD images jointly, this approach decomposes the forecasting task into sequential next-image predictions. The conditional distribution can be factorized autoregressively:
\begin{equation}
p({Y}_f \mid {Y}_l, {C}) = \prod_{t = \tau + 1}^{\tau + f} p({y}_t \mid {y}_{t-w:t-1}, {c}_{t-w:t})
\end{equation}
with temporal context window $w \leq l$.

From base distribution $p^0({y}_t) = \mathcal{N}(0, {I})$ to data distribution $p^1({y}_t)$, the conditional probability path is defined as 
$
p^{s}({y}_t \mid {z}) = \mathcal{N}\left((1 - s){y}_t^0 + s{y}_t^1, \sigma^2 {I} \right)
$ using an ODE with conditional optimal vector field $
\mu({y}_t, s \mid {z}) = {y}_t^1 - {y}_t^0
$ with $z = \{y^0
_t,y^1_t\}$, $y^0
_t \sim p^0$ and $y^1
_t \sim p^1$.

\begin{figure*}[!htbp]
    \centering
    \includegraphics[width=\linewidth]{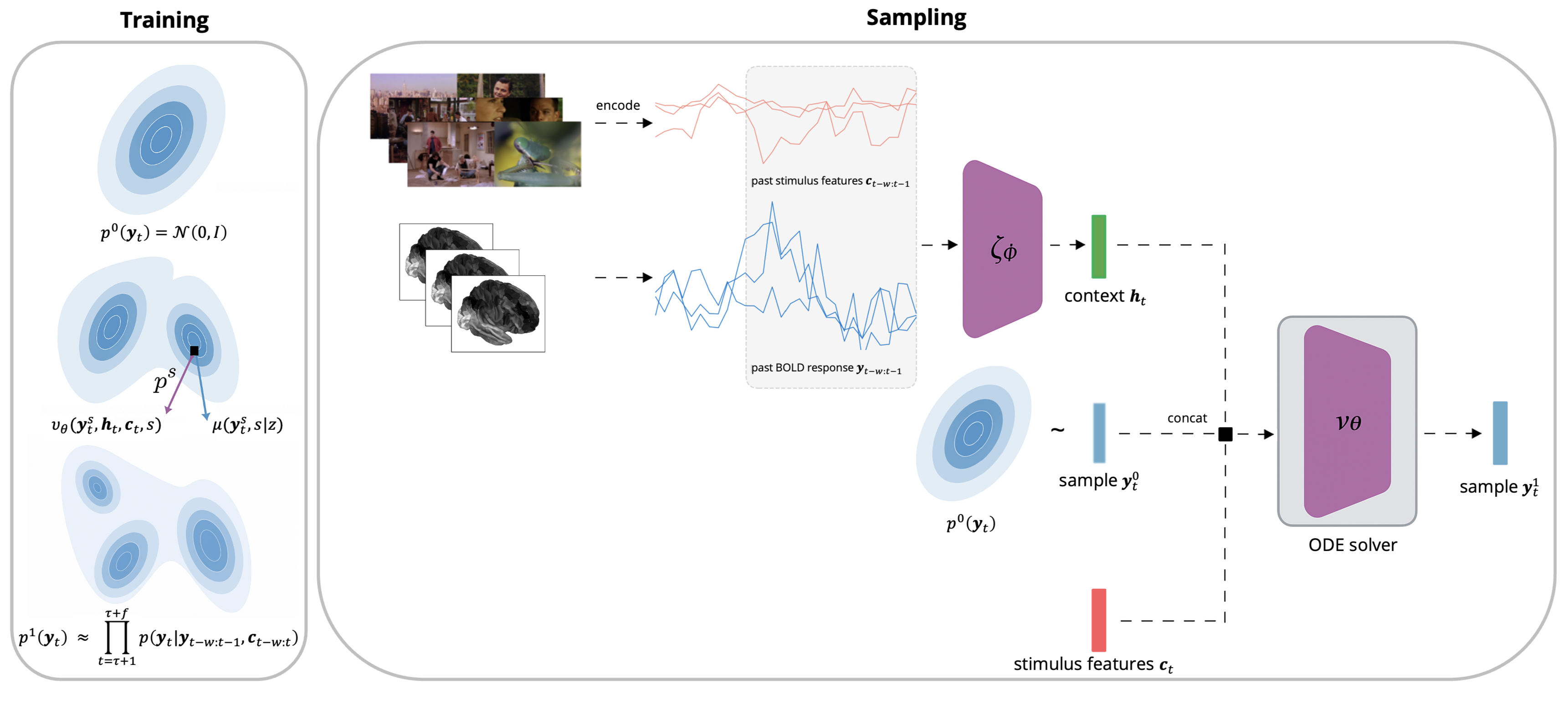}
    \caption{Autoregressive flow matching framework. Training aims to construct a probability path between a base distribution $p^0$ and the target distribution $p^1$. The path is learned by regressing a neural network $\nu_\theta$ which processes a sample from the probability path $y_t^s$ at flow step $s$, a context vector $h_t$ encoding both past BOLD images and stimulus features in a context window of length $w$, the stimulus feature $c_t$, against the target conditional velocity field $\mu$. Predictions are generated by sampling from $p^0$ and integrating $\nu_\theta$ with an ODE solver until $s=1$. Figure adapted from \textcite{elgazzar2025AFM}.}
    \label{fig:afm_fw}
\end{figure*}

The neural velocity field $\nu_\theta$ is learned by regressing against the target velocity field $\mu$ given the following training objective 
\begin{equation}
\mathcal{L}(\theta, \phi) = \mathbb{E}_{{z} \sim \pi_{0,1},\ s \sim \mathcal{U}(0,1),\ {y}_t^s \sim p^s({y}_t \mid {z})}
\left\| \mu({y}_t^s, s \mid {z}) - \nu_\theta({y}_t^s, {h}_t, {c}_t, s) \right\|^2
\end{equation}
\noindent with 
$
{h}_t = \zeta_{\phi}({y}_{t-w:t-1}, {c}_{t-w:t-1})$
being the context encoding with encoder $\zeta_{\phi}$. $\pi_{0,1}$ denotes a data coupling over pairs $({y}_t^0, {y}_t^1)$, where ${y}_t^0 \sim p^0$ and ${y}_t^1 \sim p^1$.
Sampling requires solving the ODE.
At inference, forecasts are generated in an autoregressive manner by iteratively sampling from each conditional distribution using the learned flow. Figure~\ref{fig:afm_fw} depicts a visual representation of this framework.

\paragraph{Implementation details}
Our model consists of:  
(i) an encoder $\zeta_{\phi}$ (gated recurrent units, GRU) that maps multimodal features into latent states,  
(ii) a neural dynamics model $\upsilon_{\theta}$ (time-conditioned dense layers) that evolves latent trajectories using 16-dimensional Fourier positional embeddings to encode the flow step $s$, and  
(iii) a decoder $\lambda_\theta$ (linear layer) mapping latents to predicted BOLD. An overview of selected hyperparameters can be found in \supptabref{tab:hyperparams}, details on tuning are supplied in \suppsecref{app:hyperparametertuning}.
For an evaluation of alternative encoders see \suppsecref{app:encoder_comp}.

\paragraph{Training details}
Training was performed on a single NVIDIA A100 GPU using the Adam optimizer. Each model was trained with mean squared error loss for up to 20,000 steps, with early stopping after 1,000 training steps on a 10\% validation split. The temporal context window is 30s (22 TRs). One model was trained per subject (for comparisons with group-level models see \suppsecref{app:group_vs_indiv}).

\subsection{Baselines}
As a reference to the Algonauts project 2025 challenge, we include the official general linear model (GLM) provided with the challenge toolkit with slight modifications as a baseline. Notably, this model is a simple stimulus-to-BOLD encoding approach that functions merely as a connection to the challenge and a lower-bound on performance rather than serving as a competitive baseline. 
Our main comparison is to a non-autoregressive standard flow matching (SFM) that utilizes the same historical inputs as AFM but predicts future BOLD activity segments jointly (for more information see \suppsecref{app:baselines}).

\subsection{Performance evaluation}
Performance was evaluated on the held-out \textit{Friends} season 6 test set of the Algonauts project 2025 challenge dataset. 
Predictions were averaged over 100 independent samples obtained by autoregressively sampling from the learned conditional distribution by solving the ODE with Euler's method with $dt = 0.01$ using a rolling-window approach within  the $f=10s$ (8 TRs) prediction horizon and concatenating these segments to yield full predictions. This avoids accumulation errors during inference. The first and last five samples of each episode were removed in line with the Algonauts project 2025 challenge scoring.

Performance was quantified using Pearson correlation $r$ between predicted and observed BOLD, averaged across parcels, subjects, and groups, with noise-ceiling adjustment (see \suppsecref{app:noiseceil}). 
Significance against chance-level and of model comparisons was assessed using block permutation tests~\parencite{Adolf2014}, with 30s block length, random circular shift, 10,000 repetitions and deemed statistically significant at $p < 0.05$. At the parcel level, multiple comparisons were corrected using the False Discovery Rate (FDR) procedure of Benjamini–Hochberg, while above the parcel level, Bonferroni correction was applied.

\subsection{Uncertainty quantification}
Uncertainty was evaluated using the continuous ranked probability score (CRPS), which quantifies how well the predictive distribution aligns with the observed BOLD response. For each framework, the CRPS was computed at every time step and then averaged over the full predicted time series. 

\subsection{Temporal and spatial analyses}
We conducted network-level analysis based on the seven large-scale networks defined by \textcite{yeo2011} (see \suppfigref{fig:yeonetworks}), and ablations varying context window length within the range of 0-50s and prediction horizon within the range of 0-25s.

\section{Results}\label{chap:results}

\subsection{AFM significantly outperforms baselines}
AFM significantly outperforms both GLM and non-autoregressive flow matching in forecasting future neural dynamics across all subjects (see Table~\ref{tab:mainresults_table} and \supptabref{tab:significance}). To contextualize predictive performance, we estimated noise ceilings, providing an upper bound on predictable variance given measurement noise and intrinsic variability. Across subjects, the noise ceilings are relatively low (mean $r = 0.588$; see \supptabref{tab:noiseceiling}), reflecting the inherent difficulty of predicting BOLD dynamics. 
AFM improves mean test performance by 79\% compared to GLM-based challenge baseline ($r^*_{AFM}=0.465$ vs.\ $r^*_{GLM}=0.260$) and by 11\% compared to SFM ($r^*_{AFM}=0.465$ vs.\ $r^*_{SFM}=0.420$). Although SFM attains higher train $r^*$, its larger train–test gap suggests overfitting, whereas AFM yields better generalization and earlier convergence (see \suppfigref{fig:loss_afm_sfm}).
This demonstrates that FM is a competitive approach for predicting BOLD dynamics, and that incorporating autoregressive modeling yields clear benefits. 

\begin{table*}[htbp]
\caption{Performance comparison of AFM to baselines (SFM and GLM). Noise-ceiling adjusted Pearson's correlation scores ($r^*$) per subject for GLM baseline and FM optimized individual models. Train correlations are based on season 1, test correlations are based on season 6 of the TV show \textit{Friends}. Values are rounded to three decimals. Highest values per column are highlighted in bold. All comparisons between frameworks significant given block-permutation test (10,000 repetitions, p $<$ 0.05).}
    \label{tab:mainresults_table}
        \centering
        \small
        \resizebox{\textwidth}{!}{%
        \begin{tabular}{l*{2}{ccccc}}
            \toprule
            & \multicolumn{5}{c}{Train $r^*$} & \multicolumn{5}{c}{Test $r^*$} \\
            \cmidrule(lr){2-6} \cmidrule(lr){7-11}
            Framework & mean & sub-01 & sub-02 & sub-03 & sub-05 & mean & sub-01 & sub-02 & sub-03 & sub-05\\
            \midrule
            GLM (Algonauts project 2025 challenge baseline) & 0.320 & 0.330 & 0.331 & 0.312 & 0.309 & 0.260 & 0.242 & 0.275 & 0.267 & 0.256 \\
            SFM & \textbf{0.577} & \textbf{0.597} & \textbf{0.546} & \textbf{0.584} & \textbf{0.606} & 0.423 & 0.402 & 0.405 & 0.447 & 0.438 \\
            AFM & 0.498 & 0.480 & 0.437 & 0.541 & 0.532 & \textbf{0.465} & \textbf{0.449} & \textbf{0.428} & \textbf{0.493} & \textbf{0.488}  \\
            \bottomrule
        \end{tabular}
        }

\end{table*}

AFM’s advantage was consistent across several encoder architectures (see \suppsecref{app:encoder_comp}). 
LSTM and GRU performed best, while simpler sRNNs lagged. GRU was used for all subsequent analyses due to comparable accuracy and faster training.
We also compared group-level and individual AFM models; results (see \suppsecref{app:group_vs_indiv}) show that subject-specific models consistently outperform pooled models, consistent with prior findings on inter-individual variability in fMRI responses.

\subsection{AFM achieves widespread spatial prediction performance gains}

\begin{figure*}[!htbp]
\centering
\begin{subfigure}[t]{0.32\linewidth}
    \caption{}
    \label{fig:afmres}
    \centering
    \includegraphics[width=\linewidth]{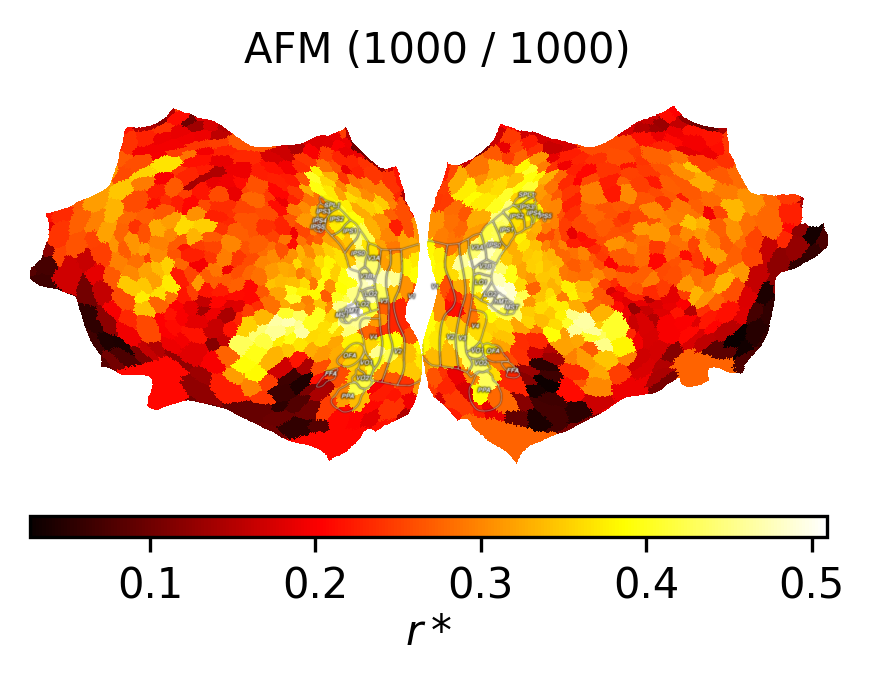}
    
\end{subfigure}
\hfill
\begin{subfigure}[t]{0.64\linewidth}
    \caption{}
    \label{fig:afmvsbaselinesflatmap}
    \centering
    \includegraphics[width=\linewidth]{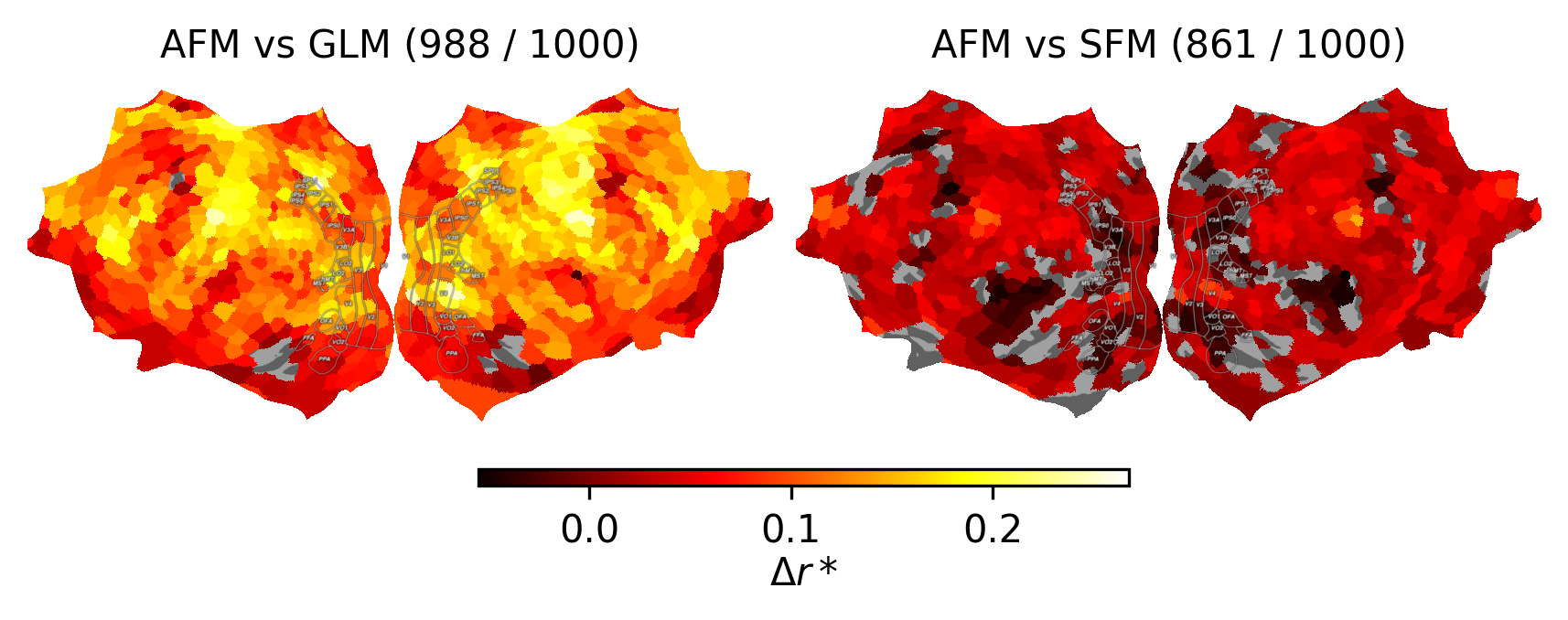}
    
\end{subfigure}

    \caption{Flatmap of mean prediction performance of AFM. A) Flatmap of parcel-wise Pearson's correlation coefficients ($r$) between predicted and observed fMRI responses averaged across subjects displayed on the FreeSurfer fsaverage brain. B) Flatmap of parcel-wise performance difference between (1) GLM and AFM and (2) SFM and AFM. Errorbar shared for difference plots.
    Insignificant parcels are depicted in grey (block-permutation test; 10,000 repetitions; $p < 0.05$, FDR-corrected across parcels). Numbers in parentheses indicate the count of significant parcels out of the total 1,000. }
\end{figure*}

Across subjects, AFM achieved widespread positive correlations throughout the cortex, with mean raw correlation scores (mean: 0.273; sub-01: 0.276; sub-02: 0.246; sub-03: 0.303; sub-05: 0.269; see Fig.~\ref{fig:afmres}), with on average all parcels significantly predicted (block-permutation test; 10,000 repetitions; $p < 0.05$; FDR correction). Spatial patterns were highly consistent across individuals, with variability largely in the magnitude rather than the location of peak correlations. The spatial distribution of prediction performance was qualitatively similar to patterns of average activity and noise ceilings, with higher correlations in parcels exhibiting greater activity and higher noise ceilings (see \suppfigref{fig:noiseceilingparcels}).

Relative to GLM, AFM increased both the number of significant parcels (1000 vs 998) and the mean correlation (mean $\Delta r^* = +0.119$; see \suppfigref{fig:glmparcelmap}). Performance difference is significant for on average 988 parcels with mean gains of $\Delta r^* = +0.120$ (see Figure~\ref{fig:afmvsbaselinesflatmap}).
Compared to SFM, AFM gains are less pronounced. Regardless, AFM outperforms mean $r^*$ (mean $\Delta r^* = +0.023$; see \suppfigref{fig:sfmparcelmap}). While AFM yields broader whole-brain performance, SFM exhibits higher maximum $r$ in few occipital and temporal parcels, indicating more localized peaks. Performance difference is significant for on average 861 parcels with mean gains of $\Delta r^* = +0.024$ (see Figure~\ref{fig:afmvsbaselinesflatmap}). 
For a detailed overview of spatial performance per subject see Supplementary Figures in \suppseconlyref{app:flatmaps}.

\subsection{AFM results in gains across large-scale cortical networks}

\begin{figure}[!htbp]
    \centering
    
    \begin{subfigure}[t]{0.343\linewidth}
    \caption{}
        
        \centering
          \raisebox{0em}{
        \includegraphics[width=\linewidth]{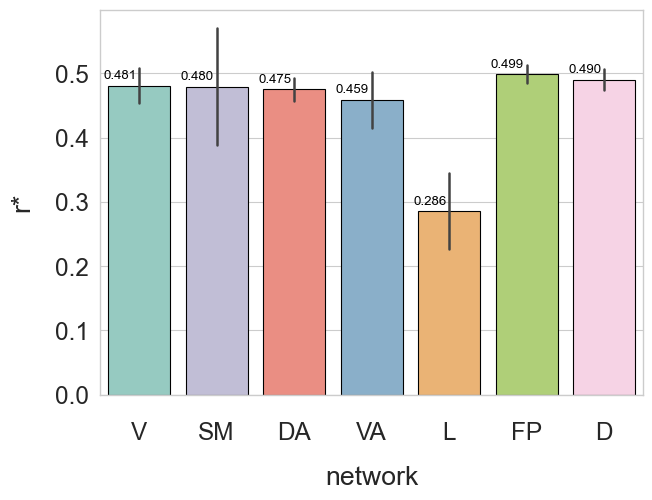}
        }
        
    \end{subfigure}
    \hfill
    \begin{subfigure}[t]{0.55\linewidth}
        \caption{}
        \label{fig:afmvsbaselinesnetworkcomp}
        \centering
        \includegraphics[width=\linewidth]{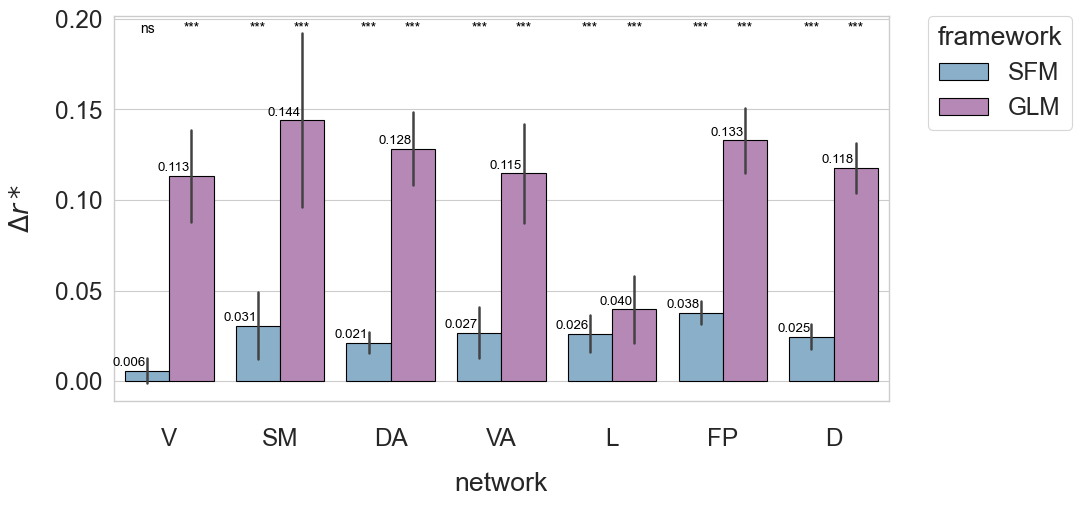}
    \end{subfigure}

    \caption{Prediction performance of optimized individual AFM models with GRU encoder across \textcite{yeo2011}'s seven brain networks. A) Noise-level-adjusted correlation test scores ($r^*$) averaged over parcels within each network. Error bars indicate variability across subjects. All results significant (block-permutation test, 10,000 repetitions, $p < 0.001$, Bonferroni correction). B) Relative performance gains ($\Delta r = r^*_{\mathrm{AFM}} - r^*_B$) of AFM over baseline frameworks (GLM, SFM). Asterisks denote significance levels of differences given a block-permutation test (10,000 repetitions, *: $p < 0.05$, **: $p < 0.001$, ***: $p < 0.001$, ****: $p < 0.0001$, ns: not significant; Bonferroni correction). Abbreviations: V: visual, SM: somatomotor, DA: dorsal attention, SV: salience / ventral attention, L: limbic, FP: frontoparietal, D: default. Results rounded to three decimals.}
     \label{fig:afmnetworkcomp}
 \end{figure}

AFM improves performance over both baselines across all seven large-scale cortical networks as specified by \textcite{yeo2011} (see Figure~\ref{fig:afmnetworkcomp} and \suppfigref{fig:yeonetworks} for a flatmap of the networks). Performance is lowest in the limbic network, consistent with its comparatively low noise ceiling (see \supptabref{tab:noiseceilingnetwork}). Gains over GLM are largest in somatomotor ($+0.144$) and frontoparietal ($+0.133$) networks and notably smallest for the limbic network ($+0.040$); gains over SFM are smaller but consistently positive ($+0.021$ to $+0.036$). 
All improvements are significant, except for gains compared to SFM in the visual network ($r^*_{\mathrm{diff}} = 0.006$, $p^* = 1.0$).

\newpage
\subsection{AFM shows comparable uncertainty quantification to SFM}

Across all subjects, AFM yields marginally improved uncertainty quantification compared to the baseline SFM. As shown in Table~\ref{tab:crps_table}, AFM achieved a lower CRPS both on average (0.369 vs. 0.370) and consistently across individual subjects, although the differences are small. For visual inspection of confidence intervals see \suppfigref{fig:uncertaintyseries}. 

\begin{table}[htpb]
    \caption{Uncertainty quantification of AFM compared to baseline SFM. Continuous ranked probability score (CRPS) per framework on test set predictions. Values are rounded to three decimals. Best values per column are highlighted in bold.}
    \label{tab:crps_table}
    \centering
    \small
    \resizebox{0.40\linewidth}{!}{%
    \begin{tabular}[t]{l*{1}{ccccc}}
            \toprule
            & \multicolumn{5}{c}{CRPS} \\
            \cmidrule(lr){2-6}
             & mean & sub-01 & sub-02 & sub-03 & sub-05\\
            \midrule
            SFM & 0.370 & 0.368 & 0.361 & 0.378 & 0.374\\
            AFM & \textbf{0.369} & \textbf{ 0.366} & \textbf{0.359 } & \textbf{0.377} & \textbf{0.373}  \\
            \bottomrule
\end{tabular}    
}
    
\end{table}

\subsection{Ablation study: AFM benefits from longer context windows and shorter prediction horizons}
\begin{figure*}[htbp]
    \centering
    
    \begin{subfigure}[t]{0.395\textwidth}
        \caption{}
        \label{fig:context_window}
        \centering
        \includegraphics[width=\linewidth]{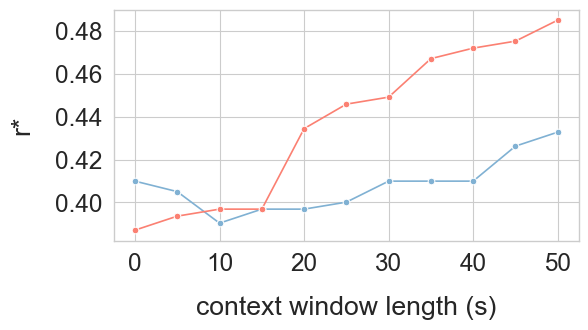}
    \end{subfigure}
      \hfill
    \begin{subfigure}[t]{0.48\textwidth}
    \caption{}
        \label{fig:prediction_window}
        \centering
        \includegraphics[width=\linewidth]{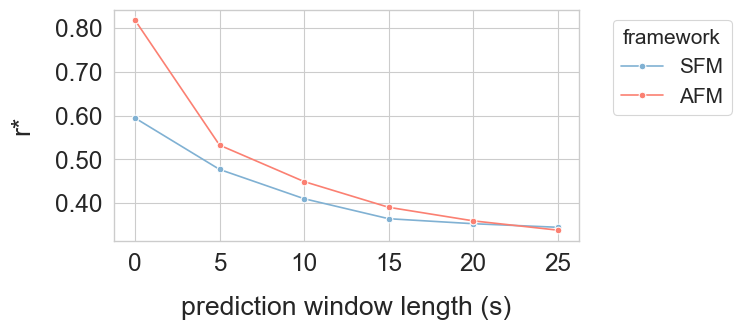}
        
    \end{subfigure}
    
    \caption{Ablation analyses of A) context window length and B) prediction window length for AFM and SFM on subject 1. Noise-ceiling adjusted correlation ($r^*$) shown for frameworks evaluated in increments of 5 (0-50s for context: 0-25 s for prediction). Markers indicate observed values.}
    \label{fig:ablation_combined}
    
\end{figure*}

We evaluated the performance of both probabilistic frameworks given different context and prediction windows on subject 1.
Increasing the context window length improved performance for both frameworks, with the effect being substantially stronger for AFM. While SFM showed only modest gains across the tested range, AFM exhibited a steady rise in $r^*$, with performance diverging from SFM once the context exceeded roughly 15s (see Figure~\ref{fig:context_window}). 

In contrast, with increased prediction window length, $r^*$ decreased for both frameworks. The decline is steeper for AFM, though AFM achieves higher $r^*$ than SFM across most window lengths, with the largest advantage at short horizons. At the longest prediction window ($s=25$), AFM's performance is slightly below that of SFM (see Figure~\ref{fig:prediction_window}).

\subsection{Ablation study: Past BOLD dynamics are critical for predictive performance}
 To isolate the contribution of past neural dynamics, we compared FM models trained with both past BOLD and stimulus features to FM models trained on only stimulus history. Removing past BOLD context resulted in a substantial reduction in predictive performance compared to the full model 
  (AFM: mean $r^*=0.465$ vs $r^*=0.284$, SFM: mean $r^*=0.423$ vs $r^*=0.162$; Table~\ref{tab:pastbold_table}). This performance drop was consistent across all subjects and reduced the model to a level only marginally above the stimulus-only official GLM baseline. 
  
\begin{table*}[!htbp]
    \caption{Performance of FM models trained using both past stimulus and BOLD dynamics compared to FM models given only stimulus history. Noise-ceiling adjusted Pearson's correlation scores ($r^*$) per subject. Test correlations are based on season 6 of the TV show \textit{Friends}. Values are rounded to three decimals.}
\label{tab:pastbold_table}
    \centering
    
    \small
    \resizebox{0.55\linewidth}{!}{%
    \begin{tabular}{l*{2}{ccccc}}
        \toprule
         & mean  & sub-01 & sub-02 & sub-03 & sub-05\\
        \midrule
        AFM (BOLD + stimulus) & \textbf{0.465} & \textbf{0.449} & \textbf{0.428} & \textbf{0.493} & \textbf{0.488} \\
        AFM (stimulus only) & 0.284 & 0.274 & 0.284 & 0.298 & 0.279 \\
        SFM (BOLD + stimulus) &  0.423 & 0.402 & 0.405 & 0.447 & 0.438 \\
        SFM (stimulus only) & 0.162 & 0.165 & 0.161 & 0.181 & 0.141 \\
        \bottomrule
    \end{tabular}
    }

\end{table*}

\section{Discussion}\label{chap:disc}
In this work, we introduced a probabilistic forecasting framework for neural dynamics based on autoregressive flow matching that produces short-term BOLD time series predictions. Using the large-scale naturalistic fMRI Algonauts project 2025 challenge dataset, we show that AFM improves predictive performance over the challenge baseline and non-autoregressive standard flow matching. Across subjects, AFM consistently achieves the highest predictive correlations, and, in comparison to SFM, reduces overfitting, with spatial and network-level analyses confirming widespread improvements across the cortex. Our findings extend prior demonstrations of AFM’s performance~\parencite{elgazzar2025AFM} to the biologically noisy and high-dimensional regime of fMRI. Our results position AFM as a promising framework for forecasting neural dynamics in naturalistic settings, as it aligns with both the temporal and stochastic properties of brain activity.

The strong performance of our probabilistic forecasting framework is largely driven by its explicit use of historical BOLD dynamics, which reflect contributions of ongoing brain dynamics, including endogenous activity, that shape future neural responses~\parencites{Amos1996, Dehaghani2025PreStimulus}.
While previous approaches to predicting future neural activity focus on encoding models, which disregard past activity, we demonstrate that utilizing past BOLD dynamics improves forecasting performance. Internal ablations show that conditioning only on stimulus features causes a sharp performance drop for both FM variants. In this stimulus-only setting, AFM effectively collapses to a one-step-ahead encoder model, achieving performance only marginally above the GLM baseline. This pattern suggests that the primary source of performance gains in both FM models arises from conditioning on past BOLD dynamics rather than requiring increased architectural complexity, as in recent challenge-winning encoding models~\parencites{tribe2025, vibe2025}, which demand substantially greater computational resources. Relating our findings to the Algonauts project 2025 challenge, we not only outperform the lower-bound reference of the simple official challenge GLM baseline encoding model, but also the more competitive contextual benchmark of the one-step ahead cluster-based multi-layer perceptron encoding model~\parencite{corsico2025islabsolutionalgonautschallenge}, which 
is, to the best of our knowledge, the only model reporting results for season 6 with a noise-ceiling adjusted score of $r^* = 0.453$. Under a comparable one-step ahead forecasting setting ablations of both our flow-matching models exceed this level ($r^* = 0.595$ for SFM; $r^* = 0.818$ for AFM), and AFM remains competitive even in our more challenging multi-step setting ($r^* = 0.465$). A direct comparison to the full Algonauts project 2025 challenge leaderboard is not possible due to differing test sets and training regimes.
Under comparable one-step-ahead settings in our setup, history-aware models outperform stimulus-only variants, suggesting that excluding past neural state can be a limiting assumption for neural dynamics forecasting.
This finding aligns with evidence that prestimulus and ongoing activity shapes sensory processing and perception~\parencites{Amos1996, Dehaghani2025PreStimulus} and motivates a shift from static stimulus–response mappings toward history-aware models.

While conditioning on past neural activity accounts for the most prominent performance gains compared to one-step ahead encoding models, our focus lies on multi-step forecasting motivated by downstream use-cases within closed-loop neurotechnology. Aligning with a dynamical systems perspective on neuroscience~\parencite{elgazzar2024UDE}, we explicitly model the temporal structure through the autoregressive formulation of AFM. This yields statistically significant and modest improvements in short-term forecasting of BOLD dynamics compared to SFM.
Across subjects, AFM outperformed SFM throughout the majority of parcels and across most major large-scale functional networks, demonstrating that autoregressive factorization provides a systematic advantage under short-horizon, context-rich conditions, that is especially valuable for naturalistic settings which engage distributed brain systems~\parencite{Sonkusare2019Naturalistic}. 
While SFM models the future trajectory jointly and does not explicitly enforce dependencies between successive future states, AFM factorizes prediction into sequential one-step conditionals, enforcing temporal continuity by autoregressive sampling in the generative process. This distinction has different implications across cortical networks. In strongly stimulus-driven regions such as the visual network~\parencite{ITO2020117141}, temporal structure is largely imposed by the stimulus itself through tight stimulus–response locking. One interpretation is that SFM can implicitly recover temporal dependencies from stimulus features alone, yielding performance comparable to AFM. In contrast, in networks whose activity is less stimulus-driven and temporally extended, such as the default and frontoparietal systems, temporal dependencies are less coupled to the stimulus~\parencite{ITO2020117141}. Here, explicitly enforcing temporal structure through autoregressive factorization provides a meaningful inductive bias, leading to improved forecasting performance.

Temporal ablations on subject 1 further suggest that increasing temporal context led to clear performance improvements for AFM across networks, whereas SFM showed only weak sensitivity to additional context. This pattern suggests that AFM is better able to translate extended history into constraints on the evolution of future activity, whereas for SFM enlarging the context window primarily increases the dimensionality of the conditioning space. AFM's context-related gains were smallest in visual cortex and larger in association networks, a pattern potentially related to previously reported hierarchical differences in intrinsic processing timescales~\parencite{Murray2014Hierarchy, Raut2020, ITO2020117141}, although these results are not fully consistent due to higher gains in the somatomotor network. However given that these analyses are limited to a single subject, the results should be interpreted cautiously and warrant further investigation (see Supplementary Figures~\suppnumref{fig:contwindow_networks} and~\suppnumref{fig:predwindow_networks}).

Moreover, by learning a simpler conditional distribution, AFM converges more rapidly during training and exhibits reduced overfitting relative to SFM. This suggests that autoregressive factorization not only improves predictive accuracy but also facilitates more stable and efficient learning, an important consideration for large-scale neural forecasting models.

Together, these analyses support the view that autoregressive formulations more effectively capture temporal dependencies in BOLD dynamics and improve forecasting performance, in line with prior work on autoregressive modeling of neural time series~\parencites{GARG2011, Deshpande2009, ROGERS2010, SETH2010, Sobczak2021}. Importantly, these gains are primarily observed in short-horizon, context-rich settings, a limitation we discuss in detail below.

Beyond predictive performance, a key strength of generative modeling for naturalistic fMRI, is their ability to quantify predictive uncertainty arising from neural variability and measurement noise as evidenced by the low noise ceiling (see \supptabref{tab:noiseceiling}) which highlights the inherent difficulty of predicting accurately. Such uncertainty estimates are crucial for risk-aware decision-making in prospective clinical settings, but also for evaluating how well a model captures the underlying neural dynamics. The slightly improved predictive distributions generated by AFM compared to SFM, which are consistently lower across subjects, networks and a majority of parcels (see Supplementary Figures~\suppnumref{fig:crps_networks} and~\suppnumref{fig:crps_diff_flatmap}), suggest that aligning the generative process with explicit temporal structure may be associated with small improvements in uncertainty calibration. Detailed statistical analyses will be needed to fully quantify this effect. Such improvements are particularly relevant given that well-calibrated predictive uncertainty is foundational for reliable model-based control and closed-loop neurotechnologies.

While these results position AFM as a promising approach for short-term probabilistic forecasting of neural dynamics, they should be interpreted in light of several limitations that also point toward directions for future work.
First, model optimization was constrained by computational resources, limiting the breadth of the search space and necessitating tuning on group-level rather than on individual models. This constraint also led us to rely on the PCA-reduced multimodal stimulus features provided by the Algonauts project 2025 challenge. However, substantial performance improvements may be achievable with more expressive features, such as those used by the challenge winner~\parencite{tribe2025}. Moreover, the performance variability we observe across encoder types within our own framework (see \suppsecref{app:encoder_comp}) suggests that further exploration of encoder architectures could yield significant additional gains.
Second, our comparison focused on each frameworks’ best-performing configuration instead of a parameter-matched evaluation, which may bias the performance comparison between AFM and SFM. Third, the sequential sampling procedure required by AFM introduces higher inference latency. This creates a trade-off between predictive accuracy and computational speed that is particularly relevant for real-time or online applications. Fourth, temporal ablations in subject 1 suggest that AFM’s advantage is largely confined to short-term forecasting and to settings with sufficiently rich temporal context. At longer prediction horizons, performance degrades more drastically than for SFM, consistent with error accumulation in autoregressive models. At shorter context windows, AFM's performance is lower than SFM's, as the latter's joint modeling can mitigate the impact of limited contextual information.

Future research should focus on the robustness and generalizability of our findings by evaluating out-of-distribution performance across stimulus types, scanners, and demographic groups~\parencite{ROHLFS2025}. Establishing such generalization is essential before deploying these models in downstream settings.
Building on this foundation, AFM’s ability to generate short-horizon, uncertainty-aware neural forecasts positions it as a promising component for emerging neurotechnological applications. Its subject-specific multi-step ahead formulation makes it particularly attractive for integration into model-predictive neural control frameworks~\parencite{Schwenzer2021}, supporting personalized closed-loop interventions for conditions such as epileptic seizures~\parencites{Chakravarthy2009, Liu2025}, and informing adaptive neurostimulation strategies~\parencite{Marta2024}.
Realizing such downstream applications will require a high level of predictive accuracy. While the prediction performance of AFM models indicates that a substantial fraction of the explainable signal is captured, the relatively low noise ceilings of this dataset (see \supptabref{tab:noiseceiling}) reflect the inherent difficulty of predicting BOLD dynamics. Consequently, even a model reaching this limit would not necessarily guarantee that the resulting forecasts are sufficient for all downstream applications, as the noisy, indirect and temporally delayed nature of the BOLD signal imposes additional challenges on neural interventions. Achieving performance suitable for such applications will therefore potentially require both richer neural measurements, such as data with a higher temporal resolution or reduced noise, and further advances in model capacity, for example through more expressive encoders or modality-informed decoders.
Finally, future methodological advances could focus on moving from the discrete-time formulation used herein toward continuous-time approaches that more faithfully capture the dynamical nature of brain activity. In particular, stochastic neural differential equations~\parencite{Kidger2022} offer a promising direction for modeling both the temporal continuity and inherent stochasticity of neural systems and can serve as a powerful basis for white-box forecasting models~\parencites{elgazzar2024LSNDE}.

\section{Conclusion}
In this work, we introduced a probabilistic, history-aware framework for forecasting neural dynamics based on autoregressive flow matching and evaluated it in a large-scale naturalistic fMRI setting. Our results show that incorporating past BOLD dynamics is a key driver of forecasting performance, substantially outperforming stimulus-only encoding approaches. Beyond this dominant effect, we find that autoregressive factorization yields consistent improvements in short-horizon forecasting relative to non-autoregressive flow matching. By enabling distributional predictions of future neural activity, our approach is suited to capture neural variability and measurement noise while further supports uncertainty-aware modeling of BOLD dynamics, a prerequisite for reliable downstream use.
While important challenges remain, our work provides a foundation for future advances in forecasting-based modeling of neural dynamics that explicitly incorporates temporal dynamics and uncertainty.

\section{Ethics}
This work uses the CNeuroMod dataset provided through the Algonauts project 2025 challenge. All participants provided informed consent for data collection and sharing. The dataset is publicly available and released under a Creative Commons CC0 license.

\section{Code availability}
The code is available in 
\href{https://github.com/nrogalla/afm-neural-dynamics.git}{this repository}.

\section{Author contributions}

NR: Methodology, Software, Formal analysis, Visualization, Writing – Original Draft.
YQ: Writing – Review \& Editing.
MS: Writing – Review \& Editing.
AE: Conceptualization, Methodology, Supervision, Writing – Review \& Editing.
MvG: Conceptualization, Methodology, Supervision, Writing – Review \& Editing.

\section{Declaration of competing interests}
The authors declare no competing interests.

\section{Acknowledgements}
This publication is part of the project Dutch Brain Interface Initiative (DBI2) with project number 024.005.022 of the research programme Gravitation which is (partly) financed by the Dutch Research Council (NWO). This project was partly funded by a fellowship supported by the European Laboratory for Learning and Intelligent Systems (ELLIS) Unit Nijmegen and the Radboud University - Maastricht University collaboration fund (Neural Control Program).

\newpage
\renewcommand{\bibname}{References}
\printbibliography

@article{Favela2021,
  author    = {Luis H. Favela},
  title     = {The dynamical renaissance in neuroscience},
  journal   = {Synthese},
  year      = {2021},
  volume    = {199},
  pages     = {2103--2127},
  doi       = {10.1007/s11229-020-02874-y}
}

@misc{meijer2024risediffusionmodelstimeseries,
      title={The Rise of Diffusion Models in Time-Series Forecasting}, 
      author={Caspar Meijer and Lydia Y. Chen},
      year={2024},
      eprint={2401.03006},
      archivePrefix={arXiv},
      primaryClass={cs.LG},
      url={https://arxiv.org/abs/2401.03006}, 
}

@misc{Kollovieh2023,
      title={Predict, Refine, Synthesize: Self-Guiding Diffusion Models for Probabilistic Time Series Forecasting}, 
      author={Marcel Kollovieh and Abdul Fatir Ansari and Michael Bohlke-Schneider and Jasper Zschiegner and Hao Wang and Yuyang Wang},
      year={2023},
      eprint={2307.11494},
      archivePrefix={arXiv},
      primaryClass={cs.LG},
      url={https://arxiv.org/abs/2307.11494}, 
}

@misc{Kidger2022,
      title={On Neural Differential Equations}, 
      author={Patrick Kidger},
      year={2022},
      eprint={2202.02435},
      archivePrefix={arXiv},
      primaryClass={cs.LG},
      url={https://arxiv.org/abs/2202.02435}
}

@misc{elgazzar2025AFM,
      title={Probabilistic Forecasting via Autoregressive Flow Matching}, 
      author={Ahmed El-Gazzar and Marcel van Gerven},
      year={2025},
      eprint={2503.10375},
      archivePrefix={arXiv},
      primaryClass={cs.LG},
      url={https://arxiv.org/abs/2503.10375}, 
}

@article{elgazzar2024UDE,
      title={Universal Differential Equations as a Common Modeling Language for Neuroscience}, 
      author={Ahmed El-Gazzar and Marcel van Gerven},
      year={2025},
      journal={Frontiers in {C}omputational {N}euroscience},
      volume={19},
      pages     = {1677930},
      doi={10.3389/fncom.2025.1677930},
    
}

@article{Ruthotto2021,
author = {Ruthotto, Lars and Haber, Eldad},
title = {An introduction to deep generative modeling},
journal = {GAMM-Mitteilungen},
volume = {44},
number = {2},
pages = {e202100008},
doi = {10.1002/gamm.202100008},
year = {2021}
}

@misc{Chen2019,
      title={Neural Ordinary Differential Equations}, 
      author={Ricky T. Q. Chen and Yulia Rubanova and Jesse Bettencourt and David Duvenaud},
      year={2019},
      eprint={1806.07366},
      archivePrefix={arXiv},
      primaryClass={cs.LG},
      url={https://arxiv.org/abs/1806.07366}, 
}

@misc{elgazzar2024LSNDE,
      title={Generative Modeling of Neural Dynamics via Latent Stochastic Differential Equations}, 
      author={Ahmed El-Gazzar and Marcel van Gerven},
      year={2024},
      eprint={2412.12112},
      archivePrefix={arXiv},
      primaryClass={q-bio.NC},
      url={https://arxiv.org/abs/2412.12112}, 
}

@misc{lipman2023,
      title={Flow Matching for Generative Modeling}, 
      author={Yaron Lipman and Ricky T. Q. Chen and Heli Ben-Hamu and Maximilian Nickel and Matt Le},
      year={2023},
      eprint={2210.02747},
      archivePrefix={arXiv},
      primaryClass={cs.LG},
      url={https://arxiv.org/abs/2210.02747}, 
}

@article{Schaefer2017,
    author = {Schaefer, Alexander and Kong, Ru and Gordon, Evan M and Laumann, Timothy O and Zuo, Xi-Nian and Holmes, Avram J and Eickhoff, Simon B and Yeo, B T Thomas},
    title = {Local-Global Parcellation of the Human Cerebral Cortex from Intrinsic Functional Connectivity {MRI}},
    journal = {Cerebral Cortex},
    volume = {28},
    number = {9},
    pages = {3095-3114},
    year = {2017},
    month = {07},
    issn = {1047-3211},
    doi = {10.1093/cercor/bhx179},
    url = {10.1093/cercor/bhx179},
    eprint = {https://academic.oup.com/cercor/article-pdf/28/9/3095/25696343/bhx179.pdf},
}

@article{Sobczak2021,
  author    = {Sobczak, Filip and He, Yujie and Sejnowski, Terrence J. and Yu, Xinxin},
  title     = {Predicting the f{MRI} Signal Fluctuation with Recurrent Neural Networks Trained on Vascular Network Dynamics},
  journal   = {Cerebral Cortex},
  year      = {2021},
  volume    = {31},
  number    = {2},
  pages     = {826--844},
  doi       = {10.1093/cercor/bhaa260},
  pmid      = {32940658},
  pmcid     = {PMC7906791},
  issn      = {1047-3211},
  
}

@ARTICLE{Adolf2014,
  
AUTHOR={Adolf, Daniela  and Weston, Snezhana  and Baecke, Sebastian  and Luchtmann, Michael  and Bernarding, Johannes  and Kropf, Siegfried },
         
TITLE={Increasing the reliability of data analysis of functional magnetic resonance imaging by applying a new blockwise permutation method},
        
JOURNAL={Frontiers in Neuroinformatics},
        
VOLUME={8},

YEAR={2014},

URL={https://www.frontiersin.org/journals/neuroinformatics/articles/10.3389/fninf.2014.00072},

DOI={10.3389/fninf.2014.00072},

ISSN={1662-5196}}

@article{Paugam2024,
    author = {Paugam, François and Pinsard, Basile and Lajoie, Guillaume and Bellec, Pierre},
    title = {A benchmark of individual auto-regressive models in a massive f{MRI} dataset},
    journal = {Imaging Neuroscience},
    volume = {2},
    pages = {1-23},
    year = {2024},
    month = {07},
    issn = {2837-6056},
    doi = {10.1162/imag_a_00228},
    url = {10.1162/imag_a_00228}
}

@misc{gifford2025,
      title={The {A}lgonauts Project 2025 Challenge: How the Human Brain Makes Sense of Multimodal Movies}, 
      author={Alessandro T. Gifford and Domenic Bersch and Marie St-Laurent and Basile Pinsard and Julie Boyle and Lune Bellec and Aude Oliva and Gemma Roig and Radoslaw M. Cichy},
      year={2025},
      eprint={2501.00504},
      archivePrefix={arXiv},
      primaryClass={q-bio.NC},
      url={https://arxiv.org/abs/2501.00504}, 
}

@INPROCEEDINGS{Feichtenhofer_2019_ICCV,
  author={Feichtenhofer, Christoph and Fan, Haoqi and Malik, Jitendra and He, Kaiming},
  booktitle={2019 IEEE/CVF International Conference on Computer Vision (ICCV)}, 
  title={SlowFast Networks for Video Recognition}, 
  year={2019},
  volume={},
  number={},
  pages={6201-6210},
  doi={10.1109/ICCV.2019.00630}}

@misc{kay2017,
      title={The Kinetics Human Action Video Dataset}, 
      author={Will Kay and Joao Carreira and Karen Simonyan and Brian Zhang and Chloe Hillier and Sudheendra Vijayanarasimhan and Fabio Viola and Tim Green and Trevor Back and Paul Natsev and Mustafa Suleyman and Andrew Zisserman},
      year={2017},
      eprint={1705.06950},
      archivePrefix={arXiv},
      primaryClass={cs.CV},
      url={https://arxiv.org/abs/1705.06950}, 
}

@inproceedings{Boyle2023,
author = {Boyle, Julie and Pinsard, Basile and Borghesani, Valentina and Paugam, Francois and DuPre, Elizabeth and Bellec, Pierre},
year = {2023},
month = {01},
pages = {},
title = {The {C}ourtois {N}euro{M}od project: quality assessment of the initial data release (2020)},
booktitle = {2023 Conference on Cognitive Computational Neuroscience}
}

@ARTICLE{Abdul2022,
  author={Abdul, Zrar Kh. and Al-Talabani, Abdulbasit K.},
  journal={IEEE Access}, 
  title={Mel Frequency Cepstral Coefficient and its Applications: A Review}, 
  year={2022},
  volume={10},
  number={},
  pages={122136-122158},
  doi={10.1109/ACCESS.2022.3223444}}

@inproceedings{devlin2019,
    title = "{BERT}: Pre-training of Deep Bidirectional Transformers for Language Understanding",
    author = "Devlin, Jacob  and
      Chang, Ming-Wei  and
      Lee, Kenton  and
      Toutanova, Kristina",
    editor = "Burstein, Jill  and
      Doran, Christy  and
      Solorio, Thamar",
    booktitle = "Proceedings of the 2019 {C}onference of the {N}orth {A}merican {C}hapter of the {A}ssociation for {C}omputational {L}inguistics: {H}uman {L}anguage {T}echnologies",
    volume = {1}, 
    month = jun,
    year = "2019",
    address = "Minneapolis, Minnesota",
    publisher = "Association for Computational Linguistics",
    url = "https://aclanthology.org/N19-1423/",
    doi = "10.18653/v1/N19-1423",
    pages = "4171--4186"
}

@InProceedings{Dvornek2019,
author="Dvornek, Nicha C.
and Li, Xiaoxiao
and Zhuang, Juntang
and Duncan, James S.",
editor="Suk, Heung-Il
and Liu, Mingxia
and Yan, Pingkun
and Lian, Chunfeng",
title="Jointly Discriminative and Generative Recurrent Neural Networks for Learning from f{MRI}",
booktitle="Machine {L}earning in {M}edical {I}maging",
volume = {11861},
year="2019",
address="Cham",
pages="382--390",
doi = {10.1007/978-3-030-32692-0_44}
}

@article{ROGERS2010,
title = {Functional {MRI} and multivariate autoregressive models},
journal = {Magnetic Resonance Imaging},
volume = {28},
number = {8},
pages = {1058-1065},
year = {2010},
issn = {0730-725X},
doi = {10.1016/j.mri.2010.03.002},
url = {https://www.sciencedirect.com/science/article/pii/S0730725X10000652},
author = {Baxter P. Rogers and Santosh B. Katwal and Victoria L. Morgan and Christopher L. Asplund and John C. Gore}
}

@article{SETH2010,
title = {A MATLAB toolbox for {G}ranger causal connectivity analysis},
journal = {Journal of Neuroscience Methods},
volume = {186},
number = {2},
pages = {262-273},
year = {2010},
issn = {0165-0270},
doi = {10.1016/j.jneumeth.2009.11.020},
url = {https://www.sciencedirect.com/science/article/pii/S0165027009006189},
author = {Anil K. Seth}
}

@article{ROEBROECK2005,
title = {Mapping directed influence over the brain using {G}ranger causality and f{MRI}},
journal = {NeuroImage},
volume = {25},
number = {1},
pages = {230-242},
year = {2005},
issn = {1053-8119},
doi = {10.1016/j.neuroimage.2004.11.017},
url = {https://www.sciencedirect.com/science/article/pii/S1053811904006688},
author = {Alard Roebroeck and Elia Formisano and Rainer Goebel}
}

@article{Deshpande2009,
author = {Deshpande, Gopikrishna and LaConte, Stephan and James, George Andrew and Peltier, Scott and Hu, Xiaoping},
title = {Multivariate {G}ranger causality analysis of f{MRI} data},
journal = {Human Brain Mapping},
volume = {30},
number = {4},
pages = {1361-1373},
doi = {10.1002/hbm.20606},
url = {https://onlinelibrary.wiley.com/doi/abs/10.1002/hbm.20606},
eprint = {https://onlinelibrary.wiley.com/doi/pdf/10.1002/hbm.20606},
year = {2009}
}

@article{yeo2011,
  title={The organization of the human cerebral cortex estimated by intrinsic functional connectivity},
  author={Yeo, B. T. T. and Krienen, F. M. and Sepulcre, J. and Sabuncu, M. R. and Lashkari, D. and Hollinshead, M. and Roffman, J. L. and Smoller, J. W. and Z{\"o}llei, L. and Polimeni, J. R. and Fischl, B. and Liu, H. and Buckner, R. L.},
  journal={Journal of Neurophysiology},
  volume={106},
  number={3},
  pages={1125--1165},
  year={2011},
  publisher={American Physiological Society},
  doi={10.1152/jn.00338.2011},
  pmid={21653723},
  pmcid={PMC3174820}
}

@misc{millidge2022,
      title={Predictive Coding: a Theoretical and Experimental Review}, 
      author={Beren Millidge and Anil Seth and Christopher L Buckley},
      year={2022},
      eprint={2107.12979},
      archivePrefix={arXiv},
      primaryClass={cs.AI},
      url={https://arxiv.org/abs/2107.12979}, 
}

@misc{hu2025,
      title={Flow{TS}: Time Series Generation via Rectified Flow}, 
      author={Yang Hu and Xiao Wang and Zezhen Ding and Lirong Wu and Huatian Zhang and Stan Z. Li and Sheng Wang and Jiheng Zhang and Ziyun Li and Tianlong Chen},
      year={2025},
      eprint={2411.07506},
      archivePrefix={arXiv},
      primaryClass={cs.LG},
      url={https://arxiv.org/abs/2411.07506}, 
}

@misc{kollovieh2025,
      title={Flow Matching with {G}aussian Process Priors for Probabilistic Time Series Forecasting}, 
      author={Marcel Kollovieh and Marten Lienen and David Lüdke and Leo Schwinn and Stephan Günnemann},
      year={2025},
      eprint={2410.03024},
      archivePrefix={arXiv},
      primaryClass={cs.LG},
      url={https://arxiv.org/abs/2410.03024}, 
}

@ARTICLE{Marta2024,
  
AUTHOR={Carè, Marta  and Chiappalone, Michela  and Cota, Vinícius Rosa },
         
TITLE={Personalized strategies of neurostimulation: from static biomarkers to dynamic closed-loop assessment of neural function},
        
JOURNAL={Frontiers in Neuroscience},
        
VOLUME={18},

YEAR={2024},

URL={https://www.frontiersin.org/journals/neuroscience/articles/10.3389/fnins.2024.1363128},

DOI={10.3389/fnins.2024.1363128},

ISSN={1662-453X}}

@ARTICLE{Liu2025,
  author={Liu, Zonglin and Qin, Yuzhen and van Gerven, Marcel and Stursberg, Olaf},
  journal={IEEE Control Systems Letters}, 
  title={Synchronization and Control in Bistable Oscillator Networks: Towards Epilepsy Regulation}, 
  year={2025},
  volume={9},
  number={},
  pages={1514-1519},
  doi={10.1109/LCSYS.2025.3582953}}

@article{Chakravarthy2009,
  author    = {Chakravarthy, N. and Sabesan, S. and Tsakalis, K. and Iasemidis, L.},
  title     = {Controlling epileptic seizures in a neural mass model},
  journal   = {Journal of Combinatorial Optimization},
  year      = {2009},
  volume    = {17},
  number    = {1},
  pages     = {98--116},
  doi       = {10.1007/s10878-008-9182-9},
  url       = {10.1007/s10878-008-9182-9}
}

@article{Schwenzer2021,
  author    = {Schwenzer, M. and Ay, M. and Bergs, T. and Abel, D.},
  title     = {Review on model predictive control: an engineering perspective},
  journal   = {The International Journal of Advanced Manufacturing Technology},
  year      = {2021},
  volume    = {117},
  number    = {5-6},
  pages     = {1327--1349},
  doi       = {10.1007/s00170-021-07682-3},
  url       = {10.1007/s00170-021-07682-3}
}

@article{Khosla2021,
author = {Meenakshi Khosla  and Gia H. Ngo  and Keith Jamison  and Amy Kuceyeski  and Mert R. Sabuncu },
title = {Cortical response to naturalistic stimuli is largely predictable with deep neural networks},
journal = {Science Advances},
volume = {7},
number = {22},
pages = {eabe7547},
year = {2021},
doi = {10.1126/sciadv.abe7547},
}

@article{ROHLFS2025,
title = {Generalization in neural networks: A broad survey},
journal = {Neurocomputing},
volume = {611},
pages = {128701},
year = {2025},
issn = {0925-2312},
doi = {10.1016/j.neucom.2024.128701},
url = {https://www.sciencedirect.com/science/article/pii/S0925231224014723},
author = {Chris Rohlfs}
}

@article{zhou2025realworld,
  author    = {Zhou, F. and Becker, B.},
  title     = {Understanding human brain function in real-world environments},
  journal   = {PLOS Biology},
  year      = {2025},
  volume    = {23},
  number    = {6},
  pages     = {e3003210},
  doi       = {10.1371/journal.pbio.3003210},
  publisher = {Public Library of Science}
}

@article{SafariMohammadbeigi2012,
  author    = {Safari, Mohammad Ali and Mohammadbeigi, M.},
  title     = {Probabilistic graphical models for effective connectivity extraction in the brain using f{MRI} data},
  journal   = {Studies in Health Technology and Informatics},
  year      = {2012},
  volume    = {180},
  pages     = {133--137}, 
doi = {10.3233/978-1-61499-101-4-133}
}

@INPROCEEDINGS{Ajith2024,
  author={Ajith, Meenu and Calhoun, Vince D.},
  booktitle={2024 IEEE EMBS International Conference on Biomedical and Health Informatics (BHI)}, 
  title={Denoising Diffusion Probabilistic Models for High-Fidelity f{MRI} Intrinsic Connectivity Network Data Generation}, 
  year={2024},
  volume={},
  number={},
  pages={1-4},
  doi={10.1109/BHI62660.2024.10913576}}

@incollection{AlowadiShenTino2016,
  author    = {Alowadi, N. and Shen, Y. and Ti{\v{n}}o, P.},
  title     = {Prototype-Based Spatio-Temporal Probabilistic Modelling of f{MRI} Data},
  booktitle = {Advances in {S}elf-{O}rganizing {M}aps and {L}earning {V}ector {Q}uantization},
  editor    = {Mer{\'e}nyi, E. and Mendenhall, M. and O'Driscoll, P.},
  series    = {Advances in Intelligent Systems and Computing},
  volume    = {428},
  publisher = {Springer},
  address   = {Cham},
  year      = {2016},
  pages     = {193--203},
  doi       = {10.1007/978-3-319-28518-4_17},
}

@incollection{LiTao2011BrainActivities,
  author       = {Li, J. and Tao, D.},
  title        = {A Probabilistic Model for Discovering High Level Brain Activities from f{MRI}},
  booktitle    = {Neural {I}nformation {P}rocessing. {ICONIP} 2011},
  editor       = {Lu, B.L. and Zhang, L. and Kwok, J.},
  volume       = {7062},
  publisher    = {Springer},
  year         = {2011},
  pages        = {329--336},
  doi          = {10.1007/978-3-642-24955-6_40},
}

@article{ZHANG2021100298,
title = {Naturalistic stimuli: A paradigm for multiscale functional characterization of the human brain},
journal = {Current Opinion in Biomedical Engineering},
volume = {19},
pages = {100298},
year = {2021},
issn = {2468-4511},
doi = {10.1016/j.cobme.2021.100298},
url = {https://www.sciencedirect.com/science/article/pii/S2468451121000386},
author = {Yizhen Zhang and Jung-Hoon Kim and David Brang and Zhongming Liu}
}

@article{SIMONY2020116461,
title = {Analysis of stimulus-induced brain dynamics during naturalistic paradigms},
journal = {NeuroImage},
volume = {216},
pages = {116461},
year = {2020},
doi = {10.1016/j.neuroimage.2019.116461},
author = {Erez Simony and Catie Chang}
}

@article{Gong2023,
  author    = {Gong, Zhengxin and Zhou, Ming and Dai, Yuxuan and Wen, Yushan and Liu, Youyi and Zhen, Zonglei},
  title     = {A large-scale f{MRI} dataset for the visual processing of naturalistic scenes},
  journal   = {Scientific Data},
  year      = {2023},
  volume    = {10},
  number    = {1},
  pages     = {559},
  doi       = {10.1038/s41597-023-02471-x}
}

@article{FRISTON2003,
title = {Dynamic causal modelling},
journal = {NeuroImage},
volume = {19},
number = {4},
pages = {1273-1302},
year = {2003},
issn = {1053-8119},
doi = {10.1016/S1053-8119(03)00202-7},
url = {https://www.sciencedirect.com/science/article/pii/S1053811903002027},
author = {K.J. Friston and L. Harrison and W. Penny}
}

@article{HARRISON2015217,
title = {Large-scale Probabilistic Functional Modes from resting state f{MRI}},
journal = {NeuroImage},
volume = {109},
pages = {217-231},
year = {2015},
issn = {1053-8119},
doi = {10.1016/j.neuroimage.2015.01.013},
author = {Samuel J. Harrison and Mark W. Woolrich and Emma C. Robinson and Matthew F. Glasser and Christian F. Beckmann and Mark Jenkinson and Stephen M. Smith}
}

@ARTICLE{Svensen2000,
  author={Svensen, M. and Kruggel, F. and von Cramon, D. Y.},
  journal={IEEE Transactions on Medical Imaging}, 
  title={Probabilistic modeling of single-trial f{MRI} data}, 
  year={2000},
  volume={19},
  number={1},
  pages={25-35},
  doi={10.1109/42.832957}}

@incollection{Battle2006,
  author    = {Battle, Alexis and Chechik, Gal and Koller, Daphne},
  title     = {Temporal and cross-subject probabilistic models for f{MRI} prediction tasks},
  booktitle = {Advances in {N}eural {I}nformation {P}rocessing {S}ystems},
  volume    = {19},
  editor    = {Sch{\"o}lkopf, Bernhard and Platt, John and Hoffman, Thomas},
  pages     = {129--136},
  publisher = {MIT Press},
  year      = {2007},
  doi       = {10.7551/mitpress/7503.003.0020},
  url       = {https://doi.org/10.7551/mitpress/7503.003.0020}
}

@ARTICLE{Li2023,
  author={Li, Wendi and Wang, Ting and Ng, Wing W. Y.},
  journal={IEEE Transactions on Neural Networks and Learning Systems}, 
  title={Population-Based Hyperparameter Tuning With Multitask Collaboration}, 
  year={2023},
  volume={34},
  number={9},
  pages={5719-5731},
  doi={10.1109/TNNLS.2021.3130896}}

@article{GARG2011,
title = {Full-brain auto-regressive modeling ({FARM}) using f{MRI}},
journal = {NeuroImage},
volume = {58},
number = {2},
pages = {416-441},
year = {2011},
issn = {1053-8119},
doi = {10.1016/j.neuroimage.2011.02.074},
url = {https://www.sciencedirect.com/science/article/pii/S1053811911002515},
author = {Rahul Garg and Guillermo A. Cecchi and A. Ravishankar Rao}
}

@misc{antoniades2024neuroformer,
      title={Neuroformer: Multimodal and Multitask Generative Pretraining for Brain Data}, 
      author={Antonis Antoniades and Yiyi Yu and Joseph Canzano and William Wang and Spencer LaVere Smith},
      year={2024},
      eprint={2311.00136},
      archivePrefix={arXiv},
      primaryClass={q-bio.NC},
      url={https://arxiv.org/abs/2311.00136}, 
}

@misc{immer2025,
      title={Forecasting Whole-Brain Neuronal Activity from Volumetric Video}, 
      author={Alexander Immer and Jan-Matthis Lueckmann and Alex Bo-Yuan Chen and Peter H. Li and Mariela D. Petkova and Nirmala A. Iyer and Aparna Dev and Gudrun Ihrke and Woohyun Park and Alyson Petruncio and Aubrey Weigel and Wyatt Korff and Florian Engert and Jeff W. Lichtman and Misha B. Ahrens and Viren Jain and Michał Januszewski},
      year={2025},
      eprint={2503.00073},
      archivePrefix={arXiv},
      primaryClass={cs.CV},
      url={https://arxiv.org/abs/2503.00073}, 
}

@misc{lueckmann2025zapbench,
      title={{ZAPB}ench: A Benchmark for Whole-Brain Activity Prediction in Zebrafish}, 
      author={Jan-Matthis Lueckmann and Alexander Immer and Alex Bo-Yuan Chen and Peter H. Li and Mariela D. Petkova and Nirmala A. Iyer and Luuk Willem Hesselink and Aparna Dev and Gudrun Ihrke and Woohyun Park and Alyson Petruncio and Aubrey Weigel and Wyatt Korff and Florian Engert and Jeff W. Lichtman and Misha B. Ahrens and Michał Januszewski and Viren Jain},
      year={2025},
      eprint={2503.02618},
      archivePrefix={arXiv},
      primaryClass={q-bio.NC},
      url={https://arxiv.org/abs/2503.02618}, 
}

@misc{lu2025,
      title={Benchmarking Probabilistic Time Series Forecasting Models on Neural Activity}, 
      author={Ziyu Lu and Anna J. Li and Alexander E. Ladd and Pascha Matveev and Aditya Deole and Eric Shea-Brown and J. Nathan Kutz and Nicholas A. Steinmetz},
      year={2025},
      eprint={2510.18037},
      archivePrefix={arXiv},
      primaryClass={cs.LG},
      url={https://arxiv.org/abs/2510.18037}, 
}

@misc{duan2025,
      title={{POCO}: Scalable Neural Forecasting through Population Conditioning}, 
      author={Yu Duan and Hamza Tahir Chaudhry and Misha B. Ahrens and Christopher D Harvey and Matthew G Perich and Karl Deisseroth and Kanaka Rajan},
      year={2025},
      eprint={2506.14957},
      archivePrefix={arXiv},
      primaryClass={q-bio.NC},
      url={https://arxiv.org/abs/2506.14957}, 
}

@article{Naselaris2011,
  author       = {Naselaris, Thomas and Kay, Kendrick N. and Nishimoto, Shinji and Gallant, Jack L.},
  title        = {Encoding and decoding in f{MRI}},
  journal      = {NeuroImage},
  year         = {2011},
  volume       = {56},
  number       = {2},
  pages        = {400--410},
  doi          = {10.1016/j.neuroimage.2010.07.073},
  issn         = {1053-8119},
  pmid         = {20691790},
  pmcid        = {PMC3037423},
  month        = {5},
}

@article{Yamins2016,
  author       = {Yamins, Daniel L. K. and DiCarlo, James J.},
  title        = {Using goal-driven deep learning models to understand sensory cortex},
  journal      = {Nature Neuroscience},
  year         = {2016},
  volume       = {19},
  number       = {3},
  pages        = {356--365},
  doi          = {10.1038/nn.4244},
}

@misc{eren2025,
      title={Multimodal Recurrent Ensembles for Predicting Brain Responses to Naturalistic Movies ({A}lgonauts 2025)}, 
      author={Semih Eren and Deniz Kucukahmetler and Nico Scherf},
      year={2025},
      eprint={2507.17897},
      archivePrefix={arXiv},
      primaryClass={q-bio.NC},
      url={https://arxiv.org/abs/2507.17897}, 
}

@misc{scholz2025,
      title={Stacked Regression using Off-the-shelf, Stimulus-tuned and Fine-tuned Neural Networks for Predicting f{MRI} Brain Responses to Movies ({A}lgonauts 2025 Report)}, 
      author={Robert Scholz and Kunal Bagga and Christine Ahrends and Carlo Alberto Barbano},
      year={2025},
      eprint={2510.06235},
      archivePrefix={arXiv},
      primaryClass={eess.IV},
      url={https://arxiv.org/abs/2510.06235}, 
}

@article{Wein2022,
    author = {Wein, S. and Schüller, A. and Tomé, A. M. and Malloni, W. M. and Greenlee, M. W. and Lang, E. W.},
    title = {Forecasting brain activity based on models of spatiotemporal brain dynamics: A comparison of graph neural network architectures},
    journal = {Network Neuroscience},
    volume = {6},
    number = {3},
    pages = {665-701},
    year = {2022},
    month = {07},
    issn = {2472-1751},
    doi = {10.1162/netn_a_00252},
    url = {10.1162/netn_a_00252},
    eprint = {https://direct.mit.edu/netn/article-pdf/6/3/665/2036019/netn_a_00252.pdf},
}

@article{Dorin2024,
  author       = {Dorin, Daniil and Kiselev, Nikita and Grabovoy, Andrey and Strijov, Vadim},
  title        = {Forecasting f{MRI} images from video sequences: linear model analysis},
  journal      = {Health Information Science and Systems},
  year         = {2024},
  volume       = {12},
  number       = {1},
  pages        = {55},
  doi          = {10.1007/s13755-024-00315-5},
  pmid         = {39554225},
  pmcid        = {PMC11568086},
}

@misc{sun2024,
      title={Predicting Human Brain States with Transformer}, 
      author={Yifei Sun and Mariano Cabezas and Jiah Lee and Chenyu Wang and Wei Zhang and Fernando Calamante and Jinglei Lv},
      year={2024},
      eprint={2412.19814},
      archivePrefix={arXiv},
      primaryClass={q-bio.NC},
      url={https://arxiv.org/abs/2412.19814}, 
}

@article {caro2024brainlm,
	author = {Caro, Josue Ortega and Fonseca, Antonio H. de O. and Averill, Christopher and Rizvi, Syed A. and Rosati, Matteo and Cross, James L. and Mittal, Prateek and Zappala, Emanuele and Levine, Daniel and Dhodapkar, Rahul M. and Han, Insu and Karbasi, Amin and Abdallah, Chadi G. and van Dijk, David},
	title = {Brain{LM}: A foundation model for brain activity recordings},
	elocation-id = {2023.09.12.557460},
	year = {2024},
	doi = {10.1101/2023.09.12.557460},
	publisher = {Cold Spring Harbor Laboratory},
	URL = {https://www.biorxiv.org/content/early/2024/01/13/2023.09.12.557460},
	eprint = {https://www.biorxiv.org/content/early/2024/01/13/2023.09.12.557460.full.pdf},
	journal = {bioRxiv}
}

@article{Guclu10005,
  author = {Güçlü, Umut and van Gerven, Marcel A. J.},
	title = {Deep Neural Networks Reveal a Gradient in the Complexity of Neural Representations across the Ventral Stream},
	volume = {35},
	number = {27},
	pages = {10005--10014},
	year = {2015},
	doi = {10.1523/JNEUROSCI.5023-14.2015},
	publisher = {Society for Neuroscience},
	issn = {0270-6474},
	URL = {https://www.jneurosci.org/content/35/27/10005},
	eprint = {https://www.jneurosci.org/content/35/27/10005.full.pdf},
	journal = {Journal of Neuroscience}
}

@article{Guclu2017,
  author       = {G{\"u}{\c c}l{\"u}, Umut and van Gerven, Marcel A. J.},
  title        = {Modeling the Dynamics of Human Brain Activity with Recurrent Neural Networks},
  journal      = {Frontiers in Computational Neuroscience},
  year         = {2017},
  volume       = {11},
  pages        = {7},
  doi          = {10.3389/fncom.2017.00007},
  pmid         = {28232797},
  pmcid        = {PMC5299026},
  month        = {2}
}

@misc{chehab2022,
      title={Deep Recurrent Encoder: A scalable end-to-end network to model brain signals}, 
      author={Omar Chehab and Alexandre Defossez and Jean-Christophe Loiseau and Alexandre Gramfort and Jean-Remi King},
      year={2022},
      eprint={2103.02339},
      archivePrefix={arXiv},
      primaryClass={q-bio.NC},
      url={https://arxiv.org/abs/2103.02339}, 
}

@inproceedings{
hu2025synthesizing,
title={Synthesizing Realistic f{MRI}: A Physiological Dynamics-Driven Hierarchical Diffusion Model for Efficient f{MRI} Acquisition},
author={Yufan Hu and Yujiang and Wuyang Li and Yixuan Yuan},
booktitle={The {T}hirteenth {I}nternational {C}onference on {L}earning {R}epresentations},
year={2025},
url={https://openreview.net/forum?id=zZ6TT254Np}
}

@inproceedings{
bayazi2024generalpurpose,
title={General-Purpose Brain Foundation Models for Time-Series Neuroimaging Data},
author={Mohammad Javad Darvishi Bayazi and Hena Ghonia and Roland Riachi and Bruno Aristimunha and Arian Khorasani and Md Rifat Arefin and Amin Darabi and Guillaume Dumas and Irina Rish},
booktitle={Neur{IPS} {W}orkshop on {T}ime {S}eries in the {A}ge of {L}arge {M}odels},
year={2024},
url={https://openreview.net/forum?id=HwDQH0r37I}
}

@inproceedings{rasul2021,
  title = 	 {Autoregressive Denoising Diffusion Models for Multivariate Probabilistic Time Series Forecasting},
  author =       {Rasul, Kashif and Seward, Calvin and Schuster, Ingmar and Vollgraf, Roland},
  booktitle = 	 {Proceedings of the 38th {I}nternational {C}onference on {M}achine {L}earning},
  pages = 	 {8857--8868},
  year = 	 {2021},
  editor = 	 {Meila, Marina and Zhang, Tong},
  volume = 	 {139},
  series = 	 {Proceedings of Machine Learning Research},
  publisher =    {PMLR},
  url = 	 {https://proceedings.mlr.press/v139/rasul21a.html}
}

@article{Feng_Miao_Zhang_Zhao_2024, title={Latent Diffusion Transformer for Probabilistic Time Series Forecasting}, volume={38}, url={https://ojs.aaai.org/index.php/AAAI/article/view/29085}, DOI={10.1609/aaai.v38i11.29085}, abstractNote={The probability prediction of multivariate time series is a notoriously challenging but practical task. This research proposes to condense high-dimensional multivariate time series forecasting into a problem of latent space time series generation, to improve the expressiveness of each timestamp and make forecasting more manageable. To solve the problem that the existing work is hard to extend to high-dimensional multivariate time series, we present a latent multivariate time series diffusion framework called Latent Diffusion Transformer (LDT), which consists of a symmetric statistics-aware autoencoder and a diffusion-based conditional generator, to implement this idea. Through careful design, the time series autoencoder can compress multivariate timestamp patterns into a concise latent representation by considering dynamic statistics. Then, the diffusion-based conditional generator is able to efficiently generate realistic multivariate timestamp values on a continuous latent space under a novel self-conditioning guidance which is modeled in a non-autoregressive way. Extensive experiments demonstrate that our model achieves state-of-the-art performance on many popular high-dimensional multivariate time series datasets.}, number={11}, journal={Proceedings of the AAAI Conference on Artificial Intelligence}, author={Feng, Shibo and Miao, Chunyan and Zhang, Zhong and Zhao, Peilin}, year={2024}, month={3}, pages={11979-11987} }

@misc{Zhang2024,
      title={Trajectory Flow Matching with Applications to Clinical Time Series Modeling}, 
      author={Xi Zhang and Yuan Pu and Yuki Kawamura and Andrew Loza and Yoshua Bengio and Dennis L. Shung and Alexander Tong},
      year={2025},
      eprint={2410.21154},
      archivePrefix={arXiv},
      primaryClass={cs.LG},
      url={https://arxiv.org/abs/2410.21154}, 
}

@article{Xiong2023,
author = {Hui Xiong  and Congying Chu  and Lingzhong Fan  and Ming Song  and Jiaqi Zhang  and Yawei Ma  and Ruonan Zheng  and Junyang Zhang  and Zhengyi Yang  and Tianzi Jiang },
title = {The Digital Twin Brain: A Bridge between Biological and Artificial Intelligence},
journal = {Intelligent Computing},
volume = {2},
number = {},
pages = {0055},
year = {2023},
doi = {10.34133/icomputing.0055},
URL = {https://spj.science.org/doi/abs/10.34133/icomputing.0055},
eprint = {https://spj.science.org/doi/pdf/10.34133/icomputing.0055}}

@article{Dehaghani2025PreStimulus,
  author    = {Dehaghani, N. S. and Zarei, M.},
  title     = {Pre-stimulus activities affect subsequent visual processing: Empirical evidence and potential neural mechanisms},
  journal   = {Brain and Behavior},
  year      = {2025},
  volume    = {15},
  number    = {2},
  pages     = {e3654},
  month     = feb,
  doi       = {10.1002/brb3.3654},
  url       = {10.1002/brb3.3654},
  pmid      = {39907172},
  pmcid     = {PMC11795279},
  publisher = {Wiley}
}

@article{Amos1996,
author = {Amos Arieli  and Alexander Sterkin  and Amiram Grinvald  and Ad Aertsen },
title = {Dynamics of Ongoing Activity: Explanation of the Large Variability in Evoked Cortical Responses},
journal = {Science},
volume = {273},
number = {5283},
pages = {1868-1871},
year = {1996},
doi = {10.1126/science.273.5283.1868},
URL = {https://www.science.org/doi/abs/10.1126/science.273.5283.1868},
eprint = {https://www.science.org/doi/pdf/10.1126/science.273.5283.1868}}

@article{Faisal2008Noise,
  title     = {Noise in the nervous system},
  author    = {Faisal, A. A. and Selen, L. P. J. and Wolpert, D. M.},
  journal   = {Nature Reviews Neuroscience},
  year      = {2008},
  volume    = {9},
  pages     = {292--303},
  doi       = {10.1038/nrn2258},
  url       = {10.1038/nrn2258}
}

@article{TOMKO1974405,
title = {Neuronal variability: non-stationary responses to identical visual stimuli},
journal = {Brain Research},
volume = {79},
number = {3},
pages = {405-418},
year = {1974},
issn = {0006-8993},
doi = {10.1016/0006-8993(74)90438-7},
url = {https://www.sciencedirect.com/science/article/pii/0006899374904387},
author = {George J. Tomko and Donald R. Crapper}
}

@article{LIU2016141,
title = {Noise contributions to the f{MRI} signal: An overview},
journal = {NeuroImage},
volume = {143},
pages = {141-151},
year = {2016},
issn = {1053-8119},
doi = {10.1016/j.neuroimage.2016.09.008},
url = {https://www.sciencedirect.com/science/article/pii/S1053811916304694},
author = {Thomas T. Liu}
}

@misc{corsico2025islabsolutionalgonautschallenge,
      title={The {ISL}ab Solution to the {A}lgonauts Challenge 2025: A Multimodal Deep Learning Approach to Brain Response Prediction}, 
      author={Andrea Corsico and Giorgia Rigamonti and Simone Zini and Luigi Celona and Paolo Napoletano},
      year={2025},
      eprint={2508.06499},
      archivePrefix={arXiv},
      primaryClass={q-bio.NC},
      url={https://arxiv.org/abs/2508.06499}, 
}

@article{Sonkusare2019Naturalistic,
  title        = {Naturalistic Stimuli in Neuroscience: Critically Acclaimed},
  author       = {Sonkusare, Saurabh and Breakspear, Michael and Guo, Christine},
  journal      = {Trends in Cognitive Sciences},
  volume       = {23},
  number       = {8},
  pages        = {699--714},
  year         = {2019},
  issn         = {1364-6613},
  doi          = {10.1016/j.tics.2019.05.004},
  url          = {https://www.sciencedirect.com/science/article/pii/S1364661319301275}
}

@misc{vibe2025,
      title={{VIBE}: Video-Input Brain Encoder for f{MRI} Response Modeling}, 
      author={Daniel Carlström Schad and Shrey Dixit and Janis Keck and Viktor Studenyak and Aleksandr Shpilevoi and Andrej Bicanski},
      year={2025},
      eprint={2507.17958},
      archivePrefix={arXiv},
      primaryClass={cs.LG},
      url={https://arxiv.org/abs/2507.17958}, 
}

@misc{tribe2025,
      title={TRIBE: TRImodal Brain Encoder for whole-brain f{MRI} response prediction}, 
      author={Stéphane d'Ascoli and Jérémy Rapin and Yohann Benchetrit and Hubert Banville and Jean-Rémi King},
      year={2025},
      eprint={2507.22229},
      archivePrefix={arXiv},
      primaryClass={cs.LG},
      url={https://arxiv.org/abs/2507.22229}, 
}

@article{ITO2020117141,
title = {A cortical hierarchy of localized and distributed processes revealed via dissociation of task activations, connectivity changes, and intrinsic timescales},
journal = {NeuroImage},
volume = {221},
pages = {117141},
year = {2020},
issn = {1053-8119},
doi = {10.1016/j.neuroimage.2020.117141},
url = {https://www.sciencedirect.com/science/article/pii/S1053811920306273},
author = {Takuya Ito and Luke J. Hearne and Michael W. Cole}
}

@article{Raut2020,
author = {Ryan V. Raut  and Abraham Z. Snyder  and Marcus E. Raichle },
title = {Hierarchical dynamics as a macroscopic organizing principle of the human brain},
journal = {Proceedings of the National Academy of Sciences},
volume = {117},
number = {34},
pages = {20890-20897},
year = {2020},
doi = {10.1073/pnas.2003383117},
URL = {https://www.pnas.org/doi/abs/10.1073/pnas.2003383117},
eprint = {https://www.pnas.org/doi/pdf/10.1073/pnas.2003383117}}

@article{Murray2014Hierarchy,
  author       = {Murray, John D. and Bernacchia, Alberto and Freedman, David J. and Romo, Ranulfo and Wallis, Jonathan D. and Cai, Xiaohui and Padoa‐Schioppa, Camillo and Pasternak, Tatiana and Seo, Hyojung and Lee, Daeyeol and Wang, Xiao-Jing},
  title        = {A hierarchy of intrinsic timescales across primate cortex},
  journal      = {Nature Neuroscience},
  volume       = {17},
  pages        = {1661--1663},
  year         = {2014},
  doi          = {10.1038/nn.3862},
  url          = {10.1038/nn.3862}
}

@article{dAscoli2026TribeV2,
  title={A foundation model of vision, audition, and language for in-silico neuroscience},
  author={d'Ascoli, St{\'e}phane and Rapin, J{\'e}r{\'e}my and Benchetrit, Yohann and Brookes, Teon and Begany, Katelyn and Raugel, Jos{\'e}phine and Banville, Hubert and King, Jean-R{\'e}mi},
  year={2026},
  howpublished = {\url{https://ai.meta.com/research/publications/a-foundation-model-of-vision-audition-and-language-for-in-silico-neuroscience/}},

}
\clearpage
\appendix
\end{document}